\documentclass[submission, Phys]{SciPost}
\pdfoutput=1
\usepackage{amsmath,amssymb,mathtools,xspace}
\usepackage{booktabs,multirow,graphicx,tabularx,slashed}
\usepackage{hyperref}
\usepackage{color,xcolor}
\usepackage[normalem]{ulem}
\usepackage{enumitem}
\usepackage{feynmp}
\usepackage{braket}
\usepackage{stackrel}
\usepackage{tikz}
\usepackage{dsfont}
\usetikzlibrary{arrows}
\usetikzlibrary{shapes.geometric}
\usepackage{float}
\usepackage{algorithm}
\usepackage[noend]{algpseudocode}

\makeatletter
\def\BState{\State\hskip-\ALG@thistlm}
\makeatother

\makeatletter
\@ifundefined{pdfoutput}{}{\DeclareGraphicsRule{*}{mps}{*}{}}
\makeatother

\makeatletter
\DeclareRobustCommand*{\bfseries}{%
   \not@math@alphabet\bfseries\mathbf
   \fontseries\bfdefault\selectfont
   \boldmath
}
\makeatother

\setitemize{itemsep=2pt,topsep=2pt,parsep=0pt,partopsep=0pt,leftmargin=*}
\setenumerate{itemsep=0pt,topsep=2pt,parsep=0pt,partopsep=0pt,labelindent=3pt,leftmargin=*}
\definecolor{Gcolor}{HTML}{3b528b}
\definecolor{Dcolor}{HTML}{e41a1c}

\tikzstyle{generator} = [rectangle, rounded corners, minimum width=3cm, minimum height=1cm,text centered, draw=Gcolor]
\tikzstyle{discriminator} = [rectangle, rounded corners, minimum width=3cm, minimum height=1cm,text centered, draw=Dcolor]
\tikzstyle{io} = [circle, trapezium left angle=70, trapezium right angle=110, minimum width=1cm, minimum height=1cm, text centered, draw=black]

\tikzstyle{process} = [rectangle, minimum width=1cm, minimum height=1cm, text centered, draw=black]
\tikzstyle{decision} = [rectangle, minimum width=1cm, minimum height=1cm, text centered, draw=black]

\tikzstyle{arrow} = [thick,->,>=stealth]
\usepackage{xcolor}

\newcommand{\mcD}{\mathcal{D}} 
\newcommand{\mcL}{\mathcal{L}}

\newcommand{\sthsy}{\sigma_{\text{th/sys}}}
\newcommand{\sth}{\sigma_{\text{th}}}
\newcommand{\sst}{\sigma_{\text{stat}}}
\newcommand{\ssy}{\sigma_{\text{syst}}}

\newcommand\one{\leavevmode\hbox{\small1\normalsize\kern-.33em1}}

\newcommand{\qqquad}{\qquad \qquad}

\def\slashchar#1{\setbox0=\hbox{$#1$}           
   \dimen0=\wd0                                 
   \setbox1=\hbox{/} \dimen1=\wd1               
   \ifdim\dimen0>\dimen1                        
      \rlap{\hbox to \dimen0{\hfil/\hfil}}      
      #1                                        
   \else                                        
      \rlap{\hbox to \dimen1{\hfil$#1$\hfil}}   
      /                                         
   \fi}

\newcommand{\madgraph}{\textsc{Madgraph}5\xspace}

\newcommand{\pytorch}{\textsc{PyTorch}\xspace}

\begin{document}

\begin{center}{\Large \textbf{
Understanding Event-Generation Networks via Uncertainties 
}}\end{center}

\begin{center}
Marco Bellagente\textsuperscript{1}, 
Manuel Hau{\ss}mann\textsuperscript{2}, 
Michel Luchmann\textsuperscript{1}, and
Tilman Plehn\textsuperscript{1}
\end{center}

\begin{center}
{\bf 1} Institut f\"ur Theoretische Physik, Universit\"at Heidelberg, Germany\\
{\bf 2} Heidelberg Collaboratory for Image Processing, Universit\"at Heidelberg, Germany
\end{center}

\begin{center}
\today
\end{center}

\section*{Abstract}
{\bf Following the growing success of generative neural networks in LHC
  simulations, the crucial question is how to control the networks
  and assign uncertainties to their event output. We show how 
  Bayesian normalizing flows or invertible networks capture uncertainties from
  the training and turn them into an uncertainty on the 
  event weight. Fundamentally, the interplay between density and
  uncertainty estimates indicates that these networks learn
  functions in analogy to parameter fits rather than binned event
  counts.}


\tikzstyle{int}=[thick,draw, minimum size=2em]

\vspace{10pt}
\noindent\rule{\textwidth}{1pt}
\tableofcontents\thispagestyle{fancy}
\noindent\rule{\textwidth}{1pt}
\vspace{10pt}

\newpage
\section{Introduction}
\label{sec:intro}

The role of first-principle simulations in our understanding of large
data sets makes LHC physics stand out in comparison to many other
areas of science. Three aspects define the application of modern big
data methods in this field:
\begin{itemize}
\item[$\cdot$] ATLAS and CMS deliver proper big data with excellent control over uncertainties;
\item[$\cdot$] perturbative quantum field theory provides consistent precision predictions;
\item[$\cdot$] fast and reliable precision simulations generate events from first principles.
\end{itemize}
The fact that experiments, field theory calculations, and simulations
control their uncertainties implies that we can work with a complete
uncertainty budget, including statistical, systematic, and theory
uncertainties. To sustain this approach at the upcoming HL-LHC, with a
data set more than 25 times the current Run~2 data set, the theory
challenge is to provide faster simulations and keep full control of
the uncertainties at the per-cent level and better.

In recent years it has been shown that modern machine learning can
improve LHC event simulations in many
ways~\cite{Butter:2020tvl}. Promising techniques include generative
adversarial networks
(GAN)~\cite{goodfellow2014generative,Creswell2018,Butter:2020qhk},
variational autoencoders~\cite{kingma2014autoencoding,Kingma2019}, and
normalizing
flows~\cite{nflow1,Kobyzev_2020,papamakarios2019normalizing,nflow_review,mller2018neural},
including invertible networks (INNs)~\cite{inn,coupling2,glow}. They
can improve phase space integration~\cite{maxim,Chen:2020nfb}, phase
space sampling~\cite{Bothmann:2020ywa,Gao:2020vdv,Gao:2020zvv}, and
amplitude computations~\cite{Bishara:2019iwh,Badger:2020uow}.  Further
developments are fully NN-based event
generation~\cite{dutch,gan_datasets,DijetGAN2,gan_phasespace,Alanazi:2020klf},
event subtraction~\cite{Butter:2019eyo}, event
unweighting~\cite{Verheyen:2020bjw,Backes:2020vka}, detector
simulation~\cite{calogan1,calogan2,fast_accurate,aachen_wgan1,aachen_wgan2,ATLASShowerGAN,ATLASsimGAN,Belayneh:2019vyx,Buhmann:2020pmy,Buhmann:2021lxj},
or parton
showering~\cite{shower,locationGAN,monkshower,juniprshower,Dohi:2020eda}.
Generative models will also improve searches for physics beyond the
Standard Model~\cite{bsm_gan}, anomaly
detection~\cite{Nachman:2020lpy,Knapp:2020dde}, detector
resolution~\cite{DiBello:2020bas,Baldi:2020hjm}, and
inference~\cite{Brehmer:2020vwc,radev2020bayesflow,Bieringer:2020tnw}. Finally,
conditional GANs and INNs allow us to invert the simulation chain to
unfold detector effects~\cite{Datta:2018mwd,fcgan} and extract the
hard scattering process at parton level~\cite{Bellagente:2020piv}. The
problem with these applications is that we know little about
\begin{enumerate}
\item how these generative networks work, and 
\item what the uncertainty on the generative network output is.
\end{enumerate}
As we will see in this paper, these two questions are closely related.

In general, we can track statistical and systematic uncertainties in
neural network outputs with Bayesian
networks~\cite{bnn_early,bnn_early2,bnn_early3,deep_errors}. Such
networks have been used in particle physics for a long
time~\cite{bnn_tev,bnn_tev2,bnn_Nu}. For the LHC we have proposed to
use them to extract uncertainties in jet
classification~\cite{Bollweg:2019skg} and jet
calibration~\cite{Kasieczka:2020vlh}. They can cover essentially all
uncertainties related to statistical, systematic, and structural
limitations of the training sample~\cite{Nachman:2019dol}.  Similar
ideas can be used as part of ensemble techniques~\cite{Araz:2021wqm}.
We propose to use a Bayesian INN (BINN) to extract uncertainties on a
generated event sample induced by the network training.

Because Bayesian networks learn the density and uncertainty maps in
one pass, their relation offers us fundamental insight into the way an
INN learns a distribution. While Bayesian
classification~\cite{Bollweg:2019skg} and regression
networks~\cite{Kasieczka:2020vlh} highlight the statistical and
systematic nature of uncertainties, our Bayesian generative network
exhibits a very different structure. We will discuss the learning
pattern of the Bayesian INN in details for a set of simple toy
processes in Sec.~\ref{sec:toy}, before we apply the network to a
semi-realistic LHC example in Sec.~\ref{sec:lhc}.

\section{Generative networks with uncertainties}
\label{sec:nets}

We start by reminding ourselves that we often assume that a generative
model has learned a phase space density perfectly, so the only
remaining source of uncertainty is the statistics of the generated
sample binned in phase space.  However, we know that such an
assumption is not realistic~\cite{Bollweg:2019skg,Kasieczka:2020vlh},
and we need to estimate the effect of statistical or systematic
limitations of the training data. The problem with such a statistical
limitation is that it is turned into a systematic shortcoming of the
generative model~\cite{gan_phasespace} --- once we generate a new
sample, the information on the training data is lost, and the only way
we might recover it is by training many networks and comparing their
outcome. For most applications this is not a realistic or economic 
option, so we will show how an alternative solution could look. 

\subsection{Uncertainties on event samples}
\label{sec:nets_unc}

Uncertainties on a simulated kinematic or phase space distribution are
crucial for any LHC analysis. For instance, we need to know to what
degree we can trust a simulated $p_T$-distribution in mono-jet searches
for dark matter. We denote the complete phase space weight for a given
phase space point as $p(x)$, such that we can illustrate a total cross
section as
\begin{align}
  \sigma_\text{tot} = \int_0^1 dx \; p(x)
  \quad \text{with} \quad p(x)>0 \; .
\end{align}
In this simplified notation $x$ stands for a generally
multi-dimensional phase space. For each phase space position, we can
also define an uncertainty $\sigma(x)$.

Two contributions to the error budget are theory and systematic
uncertainties, $\sthsy(x)$. The former reflects our ignorance of
aspects of the training data, which do not decrease when we increase
the amount of training data. The latter captures the degree to which
we trust our prediction, for instance based on self-consistency
arguments.  For example, we can account for possible large, momentum-dependent
logarithms as a simple function of phase space. 
If we use a numerical variation of the
factorization and renormalization scales to estimate a theory
uncertainty, we typically re-weight events with the scales.  Another
uncertainty arises from the statistical limitations of the training
data, $\sst(x)$. For instance in mono-jet production, the tails of the
predicted $p_T$-distribution for the Standard Model will at some point
be statistics limited. In the Gaussian limit, a statistical
uncertainty can be defined by binning the phase space and in that
limit we expect a scaling like $\sst(x) \sim \sqrt{p(x)}$, and we will
test that hypothesis in detail in Sec.~\ref{sec:toy}.

Once we know the uncertainties as a function of the phase space
position, we can account for them as additional entries in unweighted
or weighted events. For instance, relative uncertainties can be easily
added to unweighted events,
\begin{align}
  \text{ev}_i  = \begin{pmatrix} \sst/p \\ \ssy/p \\ \sth/p \\ \{ x_{\mu,j} \} \\ \{ p_{\mu,j} \} \end{pmatrix} \; ,
  \qquad \text{with $\mu=0~...~3$ for each particle $j$.} 
  \label{eq:ext_evt}
\end{align}
The entries $\sigma$ or $\sigma/p$ are smooth functions of phase
space.  The challenge in working with this definition is how to
extract $\sst$ without binning.  We will show how Bayesian networks
give us access to limited information in the training data.  Specific
theory and systematics counterparts can be either computed directly or
extracted by appropriately modifying the training
data~\cite{Bollweg:2019skg,Kasieczka:2020vlh}.

\subsection{Invertible Neural Networks}
\label{sec:nets_inn}

To model complex densities such as LHC phase space distributions, we
can employ normalizing
flows~\cite{nflow1,coupling2,glow,nflow_review}. They use the fact we
can transform a random variable $z\sim p_Z(z)$ using a bijective map
$G:z\to x$ to a random variable $x = G(z)$ with the density
\begin{align}
    p_X(x) = p_Z(z) \left|\det \frac{\partial G(z)}{\partial z}\right|^{-1} = p_Z\big(G^{-1}(x)\big)\left|\det\frac{\partial G^{-1}(x)}{\partial x}\right|\; .\label{eq:cov}
\end{align}
Given a sample $z$ from the base distribution, we can then use the map
$G$ to generate a sample from the target distribution going in the
forward direction. Alternatively, we can use a sample $x$ from the
target distribution to compute its density using the inverse
direction. We will suppress the subscripts in the distributions
$p_Z,p_X$ whenever the density is clear from the context, to lighten
the notation.

For this to be a useful approach, we require the base distribution
$p_Z$ to be simple enough to allow for efficient sample generation,
$G$ to be flexible enough for a non-trivial transformation, and its
Jacobian determinant to be efficiently computable. If these
constraints are fulfilled, $G$ gives us a powerful generative pipeline
to model the phase space density,
\begin{align}
\text{base distribution $z \sim p_Z$} 
\stackrel[\leftarrow \; \overline{G}(x)]{G(z) \rightarrow}{\xleftrightarrow{\hspace*{1.5cm}}}
\text{phase space distribution $x \sim p_X$} \;,
\label{eq:mapping}
\end{align}
where $\overline{G}(x) = G^{-1}(x)$.

To fulfill the first constraint, we choose the base distribution $p_Z$
to be a multivariate Gaussian with zero mean and an identity matrix
as the covariance.  The construction of $G$ relies on the property
that the composition of a chain of simple invertible nonlinear maps
gives us a complex map. In contrast, the determinant of the Jacobian
of the composition remains simple in the sense that we can decompose
it into the product of determinants of each of the individual
transformations.  There exists a broad literature of different
transformations, each with different strengths and
weaknesses~\cite{nflow_review}. We rely on the real non-volume
preserving flow~\cite{coupling2} in the invertible neural network
(INN) formulation~\cite{inn}.

An INN composes multiple transformation maps into coupling layers with
the following structure. The input vector $z$ into a layer is split in
half, $z = (z_1,z_2)$, allowing us to compute the output $x=(x_1,x_2)$
of the layer as
\begin{align}
\begin{pmatrix} x_1 \\ x_2 \end{pmatrix} =
\begin{pmatrix}
z_1 \odot e^{s_2(z_2)} + t_2(z_2) \\
z_2 \odot e^{s_1(x_1)} + t_1(x_1)
\end{pmatrix}\label{eq:layer1},
\end{align}
where $s_i, t_i$ ($i=1,2$) are arbitrary functions, and $\odot$ is the
element-wise product. In practice each is a small multi-layer
perceptron. This transformation has the benefit of being easily
invertible. Given a vector $x=(x_1,x_2)$ the inverse is given by
\begin{align}
\begin{pmatrix} z_1 \\ z_2 \end{pmatrix} =
\begin{pmatrix}
(x_1 - t_2(z_2)) \odot e^{-s_2(z_2)} \\
(x_2 - t_1(x_1)) \odot e^{-s_1(x_1)}
\end{pmatrix} \; .
\label{eq:layer2}
\end{align}
Additionally, its Jacobian is an upper triangular matrix
\begin{align}
\frac{\partial G(z)}{\partial z} = 
\begin{pmatrix} 
\text{diag}\left(e^{s_2(z_2)}\right) & \text{finite} \\ 
0 & \text{diag}\left(e^{s_1(x_1)}\right) 
\end{pmatrix}  \; ,
\end{align}
whose determinant is just the product of the diagonal entries,
irrespective of the entries on the off-diagonal. As such, it is
computationally inexpensive, easily composable, yet still allows for
complex transformations.

We refer to the overall map composing a sequence of such coupling
layers as $G(z;\theta)$, where we collected the parameters of the
individual nets $s$, $t$ of each layer into a joint $\theta$. Note
that each coupling layer has a separate set of nets, whose indices 
we suppress  (e.g.\ $s^l, t^l$ for the $l$-th layer).
We can then train the
overall model via a maximum likelihood approach.  It relies on the
assumption that we have access to a data set of $N$ samples
$\mathcal{D} = \{x_1,\ldots, x_N\}$ of the intractable target phase
space distribution $p_X^*(x)$ and want to fit our model distribution
$p_X(x;\theta)$ via the INN~$G$. The maximum likelihood loss is %
\begin{align}
    \mcL_\text{ML} &= - \sum_{n=1}^N \log p_X(x_n;\theta)\nonumber\\
    &=-\sum_{n=1}^N \log p_Z\big(\overline{G}(x_n;\theta)\big) + \log \left|\det \frac{\partial \overline{G}(x_n;\theta)}{\partial x_n}\right|\label{eq:MLE}\; .
\end{align}
Given the structure of $\overline{G}(x;\theta)$ and the base
distribution $p_Z$, each of the terms is tractable and can be computed
efficiently. We can approximate the sum over the complete training
data via a mini-batch and optimize the overall objective with a
stochastic gradient descent approach. Note that one can see this
maximum likelihood approach as minimizing the Kullback-Leibler (KL)
divergence between the true but unknown phase space distribution
$p_X^*(x)$ and our approximating distribution $p_X(x;\theta)$.

\subsection{Bayesian INN}
\label{sec:nets_binn}

The invertible neural net provides us with a powerful generative model
of the underlying data distribution. However, it lacks a mechanism to
account for our uncertainty in the transformation parameters $\theta$
themselves.  To model it, we switch from deterministic transformations
to probabilistic transformations, replacing the deterministic
sub-networks $s_{1,2}$ and $t_{1,2}$ in each of the coupling layers
with Bayesian neural nets. In this section, we first review the
structure of a classical Bayesian neural net
(BNN)~\cite{mackay1995probable, neal2012bayesian} as used in a
supervised learning task, and then explain how we can use BNNs for our
problem of modeling the phase space density, extending the INN into a
Bayesian invertible neural net (BINN).

\paragraph{Bayesian Neural Net}
Assuming a data set $\mcD$ consisting of $N$ pairs of observations
$(\mathbf{x}_i, y_i)$, $\mcD =\{(\mathbf{x}_1,
y_1),\ldots,(\mathbf{x}_N,y_N)\}$, in the supervised learning problem
we want to model the relation $y = f_\theta(\mathbf{x})$ through a
neural network parameterised by weights $\theta$. Placing a prior over
the weights and allowing for some observation noise, the generative
model is given as
\begin{align}
\begin{split}
    \theta &\sim p(\theta) \; , \\
    y_i|\theta, \mathbf{x}_i &\sim p(y_i|\theta, \mathbf{x}_i), \qqquad i=1,\ldots,N \; .
\end{split}
\end{align}
In case of a regression with $y_i \in \mathds{R}$ we often use a
Gaussian likelihood, $p(y_i|\theta, \mathbf{x}_i)=
\mathcal{N}\left(y_i|f_\theta(\mathbf{x}_i), \alpha^{-1}\right)$, and
a Gaussian prior over the weights $p(\theta) =
\mathcal{N}\left(\theta\vert \mathbf{0}, \beta^{-1}\mathbf{1}\right)$,
with precisions $\alpha, \beta$ and $\mathbf{1}$ the identity matrix
of suitable dimensionality~\cite{Kasieczka:2020vlh}. We are not bound
to these distributions and could for example choose a prior with a
strongly sparsifying character for further
regularization\cite{louizos2018learning,ghosh2018structured}.  Given
the highly nonlinear structure of $f_\theta$ the posterior
$p(\theta|\mcD)$ is, for practically relevant applications, analytically
intractable. While MCMC-based approaches can work for specific use
cases and small networks~\cite{springenberg2016bayesian}, they quickly
become too expensive for large architectures, so we instead rely on
variational inference (VI)~\cite{blei2017variational}. A VI-based
model approximates the posterior $p(\theta|\mcD)$ with a tractable
simplified family of distributions, $q_\phi(\theta)$, parameterized by
$\phi$. We will rely on mean-field Gaussians throughout this work,
learning a separate mean and variance parameter for each network
weight.  These parameters are learned by minimizing the KL-divergence
\begin{align}
\min_\phi \text{KL}\big(q_\phi(\theta),p(\theta|\mcD)\big)\; .
\label{eq:trueviobj}
\end{align}
However, this objective is intractable, as it relies on the unknown
posterior. Using Bayes' theorem we reformulate it as
\begin{align}
    \text{KL}\big(q_\phi(\theta),p(\theta|\mcD)\big) 
    &= -\int d\theta \; q_\phi(\theta) \log \frac{p(\mcD|\theta)p(\theta)/p(\mcD)}{q_\phi(\theta)} \notag \\
    &= -\int d\theta \; q_\phi(\theta) \log p(\mcD|\theta)  
    - \int d\theta \; q_\phi(\theta)\log \frac{p(\theta)}{q_\phi(\theta)} + \log p(\mcD)\; .
\end{align}
Now, the log evidence $\log p(\mcD)$ is bounded from below as
\begin{align}
    \log p(\mcD) 
    &= \text{KL}\big(q_\phi(\theta),p(\theta|\mcD)\big) 
    + \int d\theta \; q_\phi(\theta) \log p(\mcD|\theta)  
    - \text{KL}\big(q_\phi(\theta),p(\theta)\big) \notag \\
    &\geq \int d\theta \; q_\phi(\theta) \log p(\mcD|\theta)
    - \text{KL}\big(q_\phi(\theta),p(\theta)\big)\; .
\end{align}
Maximizing this evidence lower bound (ELBO) then is equivalent to
minimizing Eq.\eqref{eq:trueviobj}, giving us as the objective without
the intractable posterior
\begin{align}
    \mcL_\text{ELBO} = \sum_{i=1}^N \Big\langle \log p(y_i|\theta, \mathbf{x}_i)\Big\rangle_{\theta \sim q_\phi(\theta)} - \text{KL}\big(q_\phi(\theta),p(\theta)\big) \; .
\end{align}
This turns the inference problem into an optimization problem, which
allows us to take advantage of gradient descent methods such as
Adam~\cite{KingmaB14}.  As the choice of prior $p(\theta)$ is under
our control, the KL-term between the variational posterior and the
prior is tractable. The intractable expectation in the first term we
can approximate by taking $S$ samples from the variational posterior
and instead of computing the gradient over the whole data set in each
iteration switch to a stochastic gradient setup, approximating the sum
with a mini-batch of size $M$, giving us
\begin{align}
  \mcL_\text{ELBO} \approx \frac NM\sum_{i=1}^M \frac1S\sum_{s=1}^S\log p(y_i|\theta^{(s)}, \mathbf{x}_i) - \text{KL}\big(q_\phi(\theta),p(\theta)\big)
  \qquad \text{with} \quad \theta^{(s)}\sim q_\phi(\theta) \; .
\label{eq:loss_elbo}
\end{align}
In practice, it is often sufficient to approximate the expectation via
a single sample ($S=1$) per forward pass to keep the computational
cost low and further rely on local
re-parametrization~\cite{kingma2015variational} to reduce the
variance of the gradients.

\paragraph{Bayesian INN}
As discussed in Sec.~\ref{sec:nets_inn}, our generative model of the
density consists of a map $G: z \to x$ from a base distribution
$p_Z(z)$ to the phase-space $p_X(x)$ parameterized via an
INN. Replacing the deterministic sub-networks $s_{1,2}$ and $t_{1,2}$
in Eq.\eqref{eq:layer1} with BNNs, we get as the generative pipeline
for our BINN
\begin{align}
\begin{split}
    \theta &\sim p(\theta),\\ 
    x|\theta &\sim p_X(x|\theta)= p_Z(\overline{G}(x;\theta))\Big|\det \frac{\partial \overline{G}(x;\theta)}{\partial x}\Big|\; .
\end{split}
\end{align}
Given a set of $N$ observations $\mcD = \{x_1,\ldots,x_N\}$, we can
approximate the intractable posterior $p(\theta|\mcD)$ as before with
a mean-field Gaussian as the variational posterior
$q_\phi(\theta)$. Learning the map and the posterior then is achieved
by maximizing the equivalent of the ELBO loss in
Eq.\eqref{eq:loss_elbo} for event samples,
\begin{align}
   \mcL &= \sum_{n=1}^N \langle \log p_X(x_n|\theta)\rangle_{\theta\sim q_\phi(\theta)} - \text{KL}\big(q_\phi(\theta), p(\theta)\big)\nonumber \\
   &= \sum_{n=1}^N \Big\langle \log p_Z\big(\overline{G}(x_n;\theta)\big) + \log \Big|\det \frac{\partial \overline{G}(x_n;\theta)}{\partial x_n}\Big|\Big\rangle_{\theta\sim q_\phi(\theta)} - \text{KL}\big(q_\phi(\theta), p(\theta)\big)\nonumber \\
   &\approx \frac NM  \sum_{m=1}^M \frac1S \sum_{s=1}^S\log p_Z\big(\overline{G}(x_m;\theta^{(s)})\big) + \log \Big|\det \frac{\partial \overline{G}(x_m;\theta^{(s)})}{\partial x_m}\Big| - \text{KL}\big(q_\phi(\theta), p(\theta)\big)\;,
\end{align}
with a mini-batch of size $M$ and $S$ samples $\theta^{(s)}$ from the
variational posterior $q_\phi(\theta)$. By design all three terms, the
log likelihood, the log determinant of the Jacobian as well as the
Kullback-Leibler divergence can be computed easily. Automatic
differentiation~\cite{pytorch} allows us to get the gradients of
$\mcL$ with respect to $\phi$ in order to fit our generative pipeline
via a stochastic gradient descent update scheme.

\section{Toy events with uncertainties}
\label{sec:toy}

\begin{table}[b!]
\begin{small} \begin{center}
\begin{tabular}{l r }
\toprule
Parameter & Flow\\
\midrule
Hidden layers (per block) & 3 \\
Units per hidden layer & 32 \\
Batch size & 512\\
Epochs & 300 \\
Trainable weights &  75k  \\
Optimizer & Adam \\
($\alpha$, $\beta_1$, $\beta_2$)  & ($1\times10^{-3}$, 0.9, 0.999) \\
Coupling layers & 20 \\
Training size & 300k \\
Prior width & 1 \\
\bottomrule
\end{tabular}
\end{center} \end{small}
\caption{Hyper-parameters for all toy models, implemented in \pytorch(v1.4.0)~\cite{pytorch}.}
\label{tab:toy_params}
\end{table}

Before we tackle a semi-realistic LHC setup, we first study the
behavior of BINNs for a set of toy examples, namely distributions over
the minimally allowed two-dimensional parameter space where in one
dimension the density is flat. Aside from the fact that these toy
examples illustrate that the BINN actually constructs a meaningful
uncertainty distribution, we will use the combination of density and
uncertainty maps to analyse how an INN actually learns a density
distributions. We will see that the INN describes the density map in
the sense of a few-parameter fit, rather than numerically encoding
patches over the parameter space independently.

The default architecture for our toy models is a
network with 32 units per layer, three layers per
coupling block, and a total of 20 coupling blocks. It's implemented in \textsc{PyTorch}~\cite{pytorch}. More details are
given in Tab.~\ref{tab:toy_params}.  The most relevant hyperparameter
is the number of coupling blocks in that more blocks provide a more
stable performance with respect to several trainings of the same
architecture.  Generally, moderate changes for instance of the number
of units per layer do not have a visible impact on the
performance. For each of the trainings we use a sample of 300k
events. The widths of the Gaussian priors is set to one. We check that
variations of this over several orders of magnitude did not have a
significant impact on the performance.

\subsection{Wedge ramp}
\label{sec:toy_wedge}

\begin{figure}[b!]
\centering
\includegraphics[width=0.32\textwidth, page=1]{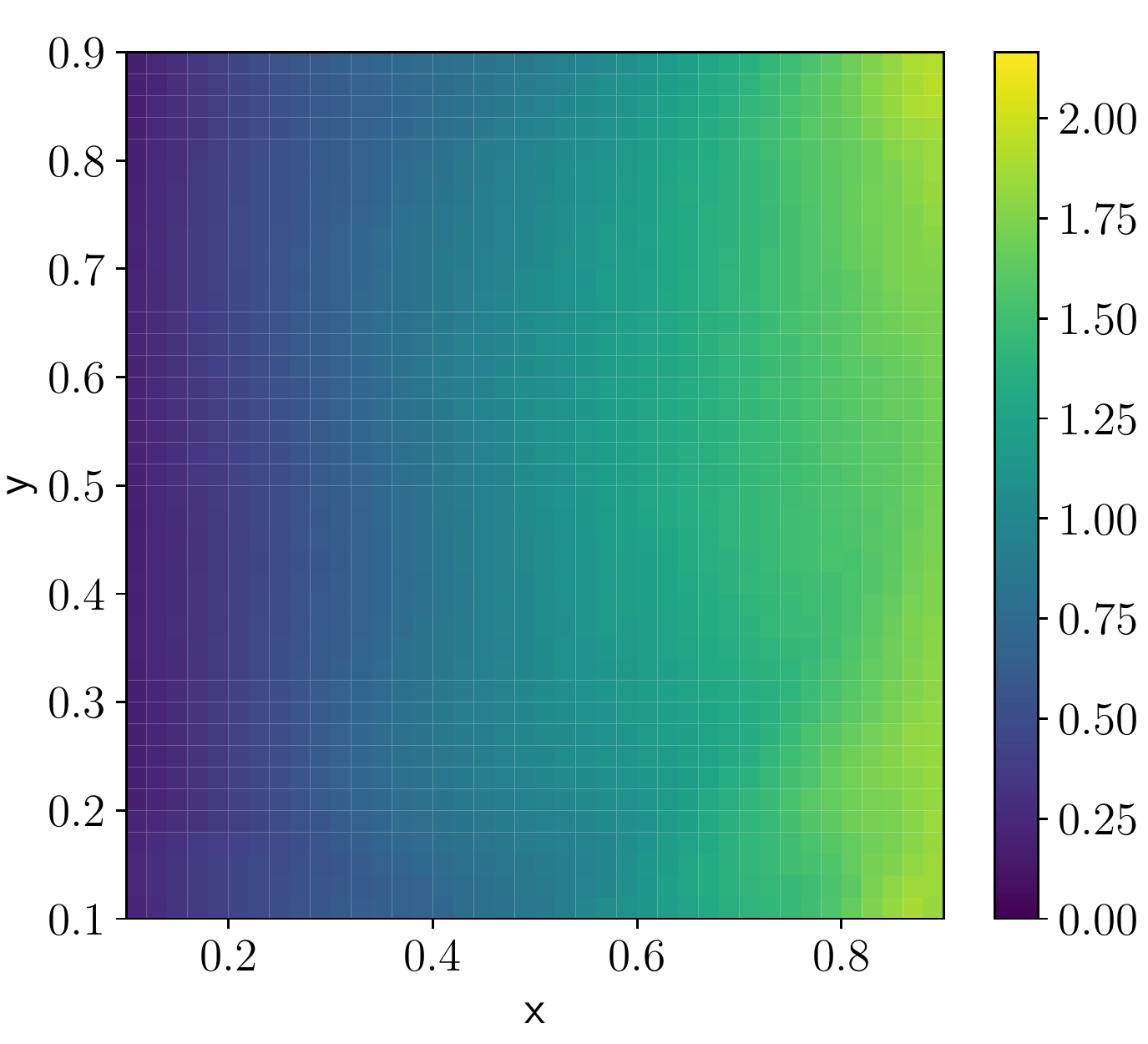}
\includegraphics[width=0.32\textwidth, page=1]{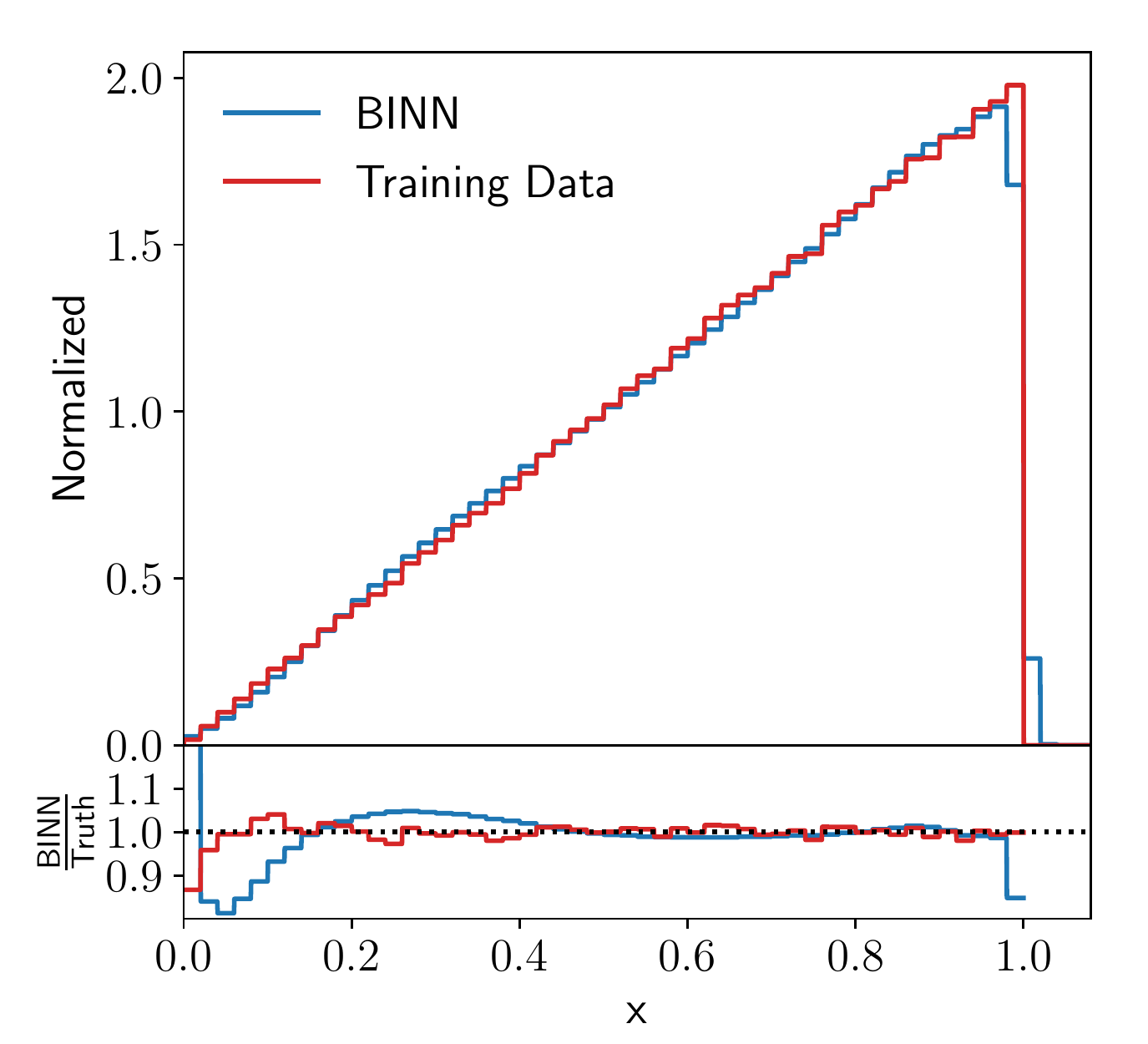}
\includegraphics[width=0.32\textwidth, page=2]{./figs/new_figures/linear_1dhists}
\caption{Two-dimensional and marginal densities for the linear wedge
  ramp.}
\label{fig:linear_ring_hists}
\end{figure}

Our first toy example is a two-dimensional ramp distribution, linear
in one direction and flat in the other,
\begin{align}
p(x, y) = \text{Linear}(x \in [0, 1]) \times \text{Const}(y \in [0, 1]) = x \times 2 \; . 
\label{eq:linear_dens}
\end{align}
The second term ensures that the distribution $p(x,y)$ is normalized
to one, and the network output is shown in
Fig.~\ref{fig:linear_ring_hists}. The network output consists of unweighted
events in the two-dimensional parameters space, $(x,y)$. We show
one-dimensional distributions after marginalizing over the unobserved
direction and find that the network reproduces
Eq.\eqref{eq:linear_dens} well.

\begin{figure}[t]
\centering
\includegraphics[width=0.32\textwidth, page=1]{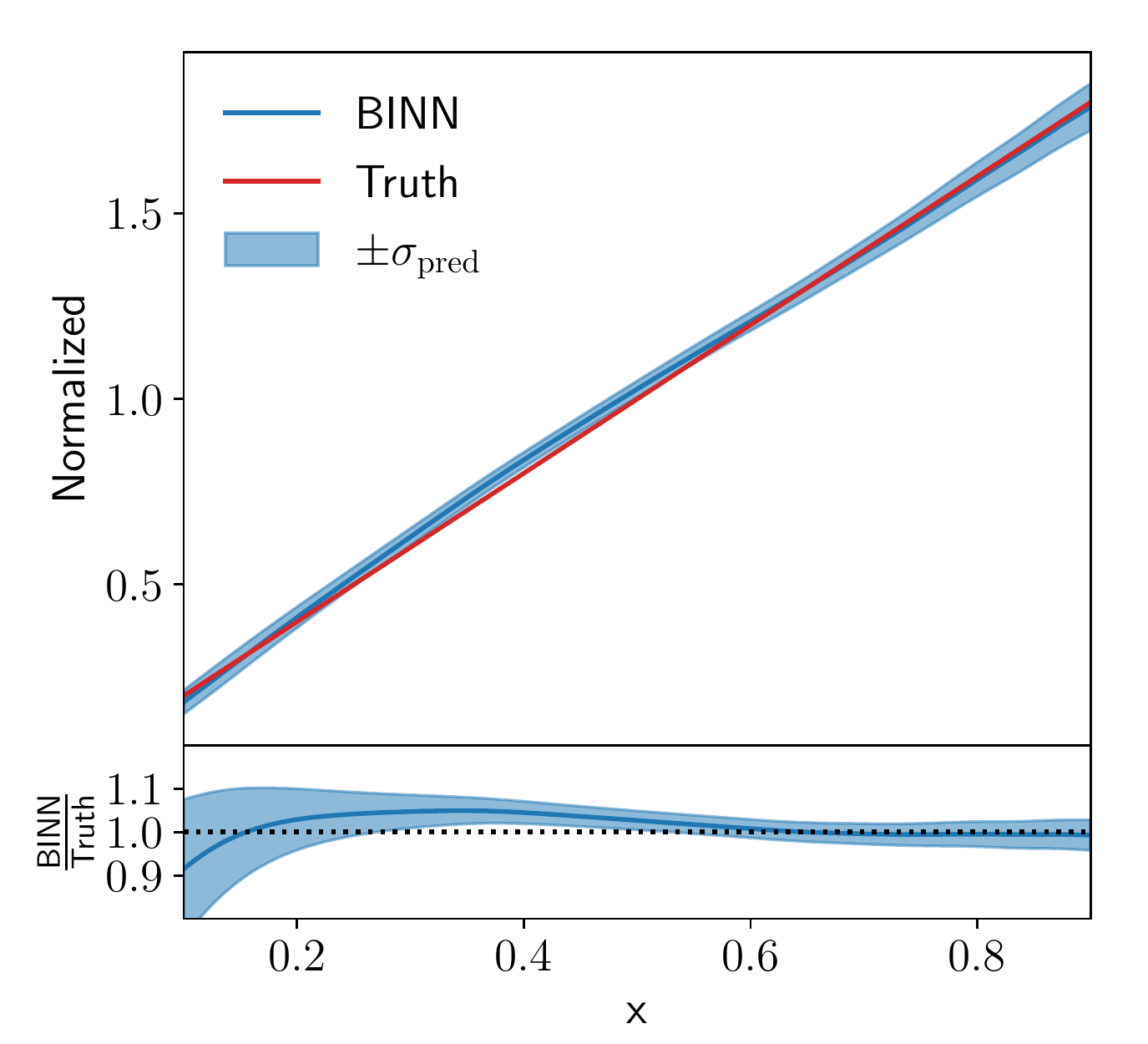}
\includegraphics[width=0.32\textwidth, page=2]{./figs/new_figures/linear_1dplots}
\includegraphics[width=0.32\textwidth, page=3]{./figs/new_figures/linear_1dplots}
\caption{Density and predictive uncertainty distribution for the wedge
  ramp. In the left panel the density and uncertainty are averaged
  over several lines with constant $y$. In the central and right
  panels, the uncertainty band on $\sigma_\text{pred}$ is given by
  their variation.  The green curve represents a two-parameter fit to
  Eq.\eqref{eq:fit_wedge}.}
  \label{fig:linear_unc}
\end{figure}

In Fig.~\ref{fig:linear_unc} we include the predictive uncertainty
given by the BINN. For this purpose we train a network on the
two-dimensional parameter space and evaluate it for a set of points
with $x \in [0,1]$ and a constant $y$-value. In the left panel we
indicate the predictive uncertainty as an error bar around the density
estimate. Throughout the paper we always remove the phase space
boundaries, because we know that the network is unstable there, and
the uncertainties explode just like we expect. For this example, this is taken into account by restricting $x, y \in [0.1,0.9]$. The relative
uncertainty grows for small values of $x$ and hence small values of
$p(x,y)$, and it covers the deviation of the extracted density from
the true density well. These features are common to all our network
trainings. In the central and right panel of Fig.~\ref{fig:linear_unc}
we show the relative and absolute predictive uncertainties. The error
bar indicates how much $\sigma_\text{pred}$ varies for different
choices of $y$. We compute it as the standard deviation of different
values of $\sigma_\text{pred}$, after confirming that the central
values agree within this range. As expected, the relative uncertainty
decreases towards larger $x$. However, the absolute uncertainty shows
a distinctive minimum in $\sigma_\text{pred}$ around $x \approx
0.45$. This minimum is a common feature in all our trainings, so we
need to explain it.

To understand this non-trivial uncertainty distribution
$\sigma_\text{pred}(x)$ we focus on the non-trivial $x$-coordinate and
its linear behavior
\begin{align}
  p(x) = a  x + b
  \qquad \text{with} \qquad x \in [0,1] \; .
\end{align}
Because the network learns a density, we can remove $b$ by
fixing the normalization,
\begin{align}
  p(x) = a \left( x - \frac{1}{2} \right) + 1 \; .
\end{align}
If we now assume that a network acts like a fit of $a$, we can relate
the uncertainty $\Delta a$ to an uncertainty in the density,
\begin{align}
\sigma_\text{pred} \equiv \Delta p \approx \left| x - \frac{1}{2} \right| \; \Delta a \; .
\label{eq:simple_wedge}
\end{align}
The absolute value appears because the uncertainties are defined to be
positive, as encoded in the usual quadratic error propagation. The
uncertainty distribution has a minimum at $x=1/2$, close to the
observed value in Fig.~\ref{fig:linear_unc}.

The differences between the simple prediction in
Eq.\eqref{eq:simple_wedge} and our numerical findings in
Fig.~\ref{fig:linear_unc} is that the predictive uncertainty is not
symmetric and does not reach zero. To account for these sub-leading
effects we can expand our very simple ansatz to
\begin{align}
  p(x) = a  x + b
  \qquad \text{with} \qquad x \in [x_\text{min},x_\text{max}] \; .
\label{eq:fund_wedge}
\end{align}
Using the normalization condition we again remove $b$ and find
\begin{align}
  p(x)
  = a x
  +  \frac{ 1 - \frac{a}{2}(x_\text{max}^2 - x_\text{min}^2) }{ x_\text{max} - x_\text{min} } \; .
\end{align}
Again assuming a fit-like behavior of the flow network we expect for
the predictive uncertainty
\begin{align}
\sigma_\text{pred}^2 \equiv (\Delta p)^2 =
    \left( x - \frac{1}{2} \right)^2 (\Delta a)^2
    + \left(1 + \frac{a}{2} \right)^2 (\Delta x_\text{max} )^2
    + \left(1 - \frac{a}{2} \right)^2 (\Delta x_\text{min} )^2 \; .
\label{eq:fit_wedge}
\end{align}
Adding $x_\text{min}$ or $x_\text{max}$ leads to an $x$-independent offset
and does not change the $x$-dependence of the predictive
uncertainty. The slight shift of the minimum and the asymmetry between
the lower and upper boundaries in $x$ are not explained by this
argument.  We ascribe them to boundary effects, specifically the
challenge for the network to describe the correct approach towards
$p(x) \to 0$.

The green line in Fig.~\ref{fig:linear_unc} gives
a two-parameter fit of $\Delta a$ and $\Delta x_\text{max}$ to the
$\sigma_\text{pred}$ distribution from the BINN. It indicates that
there is a hierarchy in the way the network extracts the
$x$-independent term with high precision, whereas the uncertainty on
the slope $a$ is around 4\%.

\subsection{Kicker ramp}
\label{sec:toy_kicker}

\begin{figure}[b!]
\centering
\includegraphics[width=0.32\textwidth, page=1]{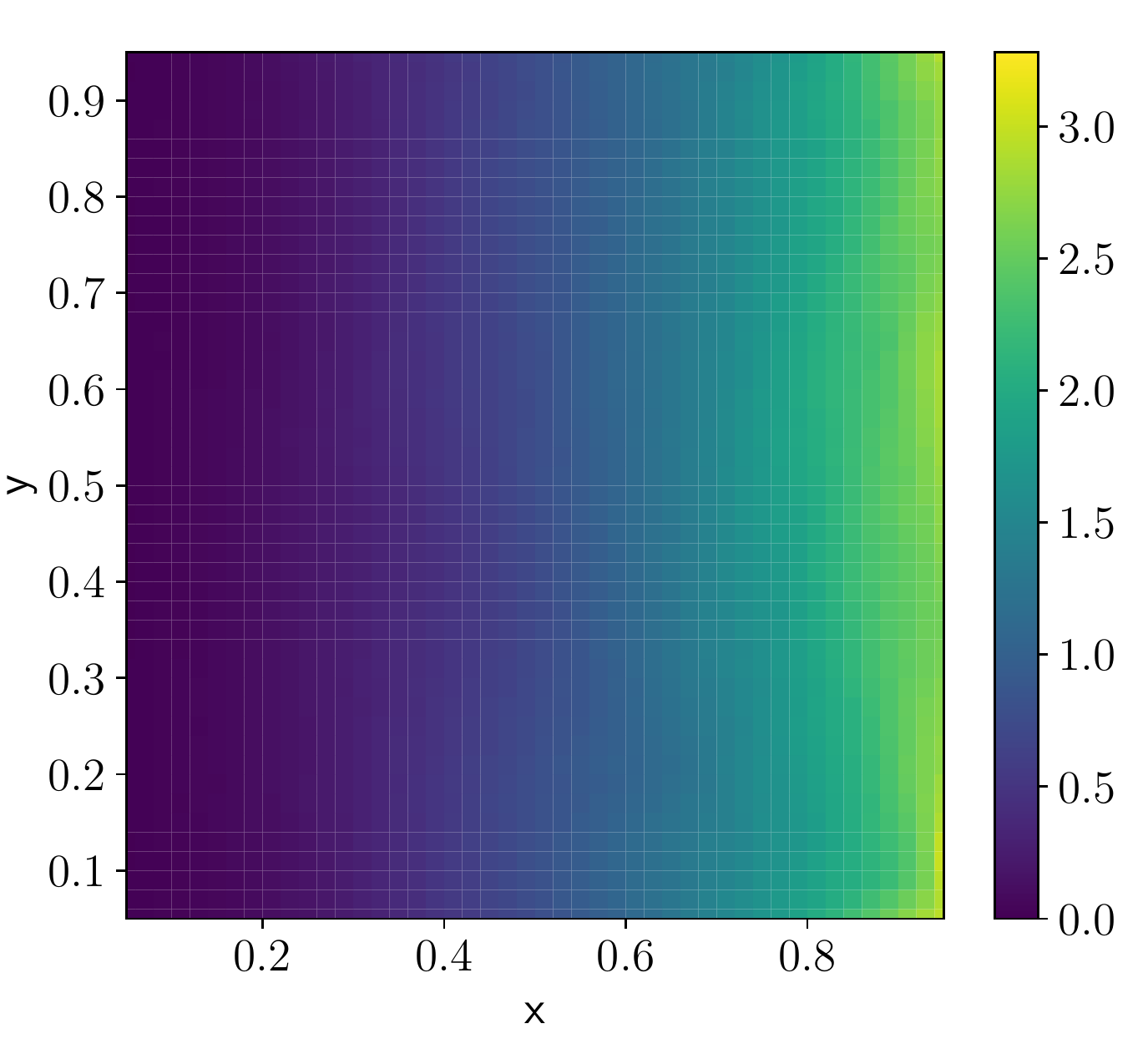}
\includegraphics[width=0.32\textwidth, page=1]{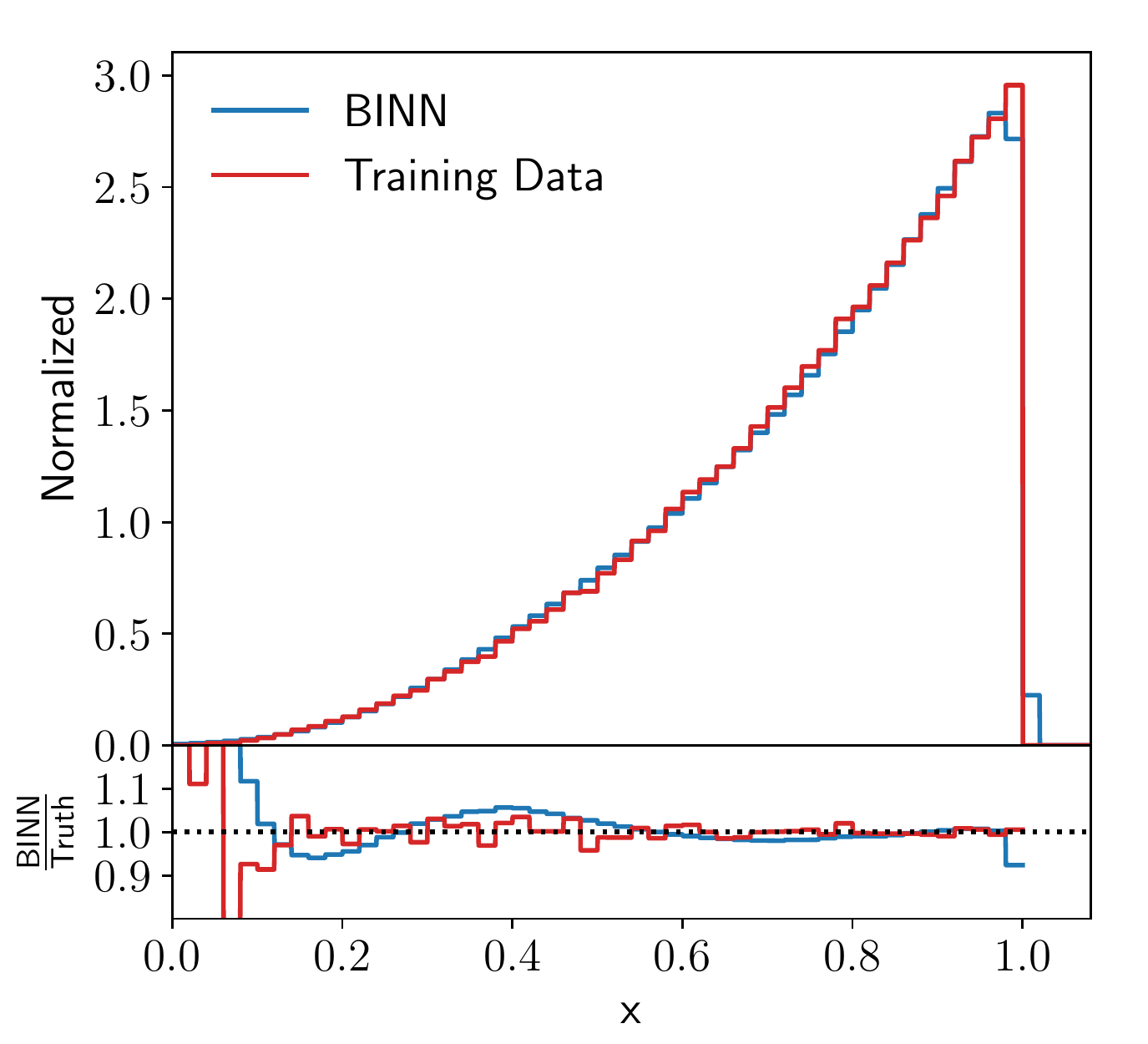}
\includegraphics[width=0.32\textwidth, page=2]{./figs/new_figures/quadratic_1dhists}
\caption{Two-dimensional and marginal densities for the quadratic
  kicker ramp.}
\label{fig:quadratic_hists}
\end{figure}

We can test our findings from the linear wedge ramp using the slightly
more complex quadratic or kicker ramp,
\begin{align}
  p(x, y) =  \text{Quadr} (x\in[0,1]) \times \text{Const} (y \in[0, 1])
  = x^2 \times 3 \; .
\label{eq:quadratic_dens}
\end{align}
We show the results from the network training for the density in
Fig.~\ref{fig:quadratic_hists} and find that the network describes the
density well, limited largely by the flat, low-statistics approach
towards the lower boundary with $p(x) \to 0$.

\begin{figure}[t]
\centering
\includegraphics[width=0.32\textwidth, page=1]{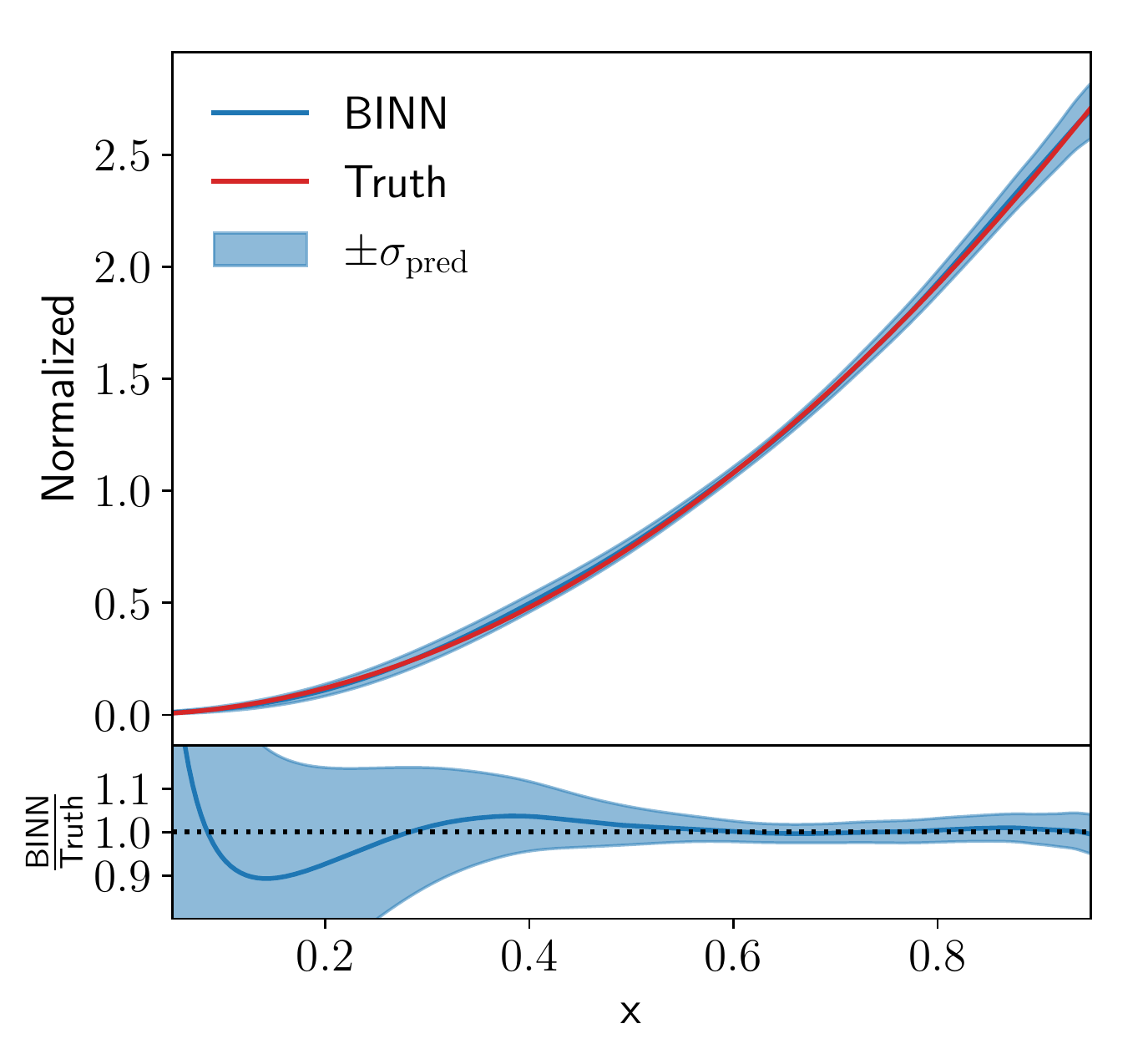}
\includegraphics[width=0.32\textwidth, page=2]{./figs/new_figures/quadratic_1dplots}
\includegraphics[width=0.32\textwidth, page=3]{./figs/new_figures/quadratic_1dplots}
\caption{Density and predictive uncertainty distribution for the
  kicker ramp. In the left panel the density and uncertainty are
  averaged over several lines with constant $y$. In the central and
  right panels, the uncertainty band on $\sigma_\text{pred}$ is given
  by their variation.  The green curve represents a two-parameter fit
  to Eq.\eqref{eq:fit_kicker}.}
\label{fig:quadratic_unc}
\end{figure}

In complete analogy to Fig.~\ref{fig:linear_unc} we show the complete
BINN output with the density $p(x,y)$ and the predictive uncertainty
$\sigma_\text{pred}(x,y)$ in Fig.~\ref{fig:quadratic_unc}. As for the
linear case, the BINN reproduces the density well, deviations from the
truth being within the predictive uncertainty in all points of phase
space. We remove the phase space boundaries restricting $x, y \in [0.05, 0.95]$, as the network becomes unstable and the predictive uncertainties grows correspondingly.  The
indicated error bar on $\sigma_\text{pred}(x,y)$ is given by the
variation of the predictions for different $y$-values, after ensuring
that their central values agree.  The relative uncertainty at the
lower boundary $x = 0$ is large, reflecting the statistical limitation
of this phase space region. An interesting feature appears again in
the absolute uncertainty, namely a maximum-minimum combination as a
function of $x$.

Again in analogy to Eq.\eqref{eq:fund_wedge} for the wedge ramp, we
start with the parametrization of the density
\begin{align}
  p(x) = a \, (x - x_0)^2
  \qquad \text{with} \qquad x \in [x_0, x_\text{max}] \; ,
\end{align}
where we assume that the lower boundary coincides with the minimum and
there is no constant offset. We choose to describe this density
through the minimum position $x_0$, coinciding the the lower end of
the $x$-range, and $x_\text{max}$ as the second parameter. The
parameter $a$ can be eliminated through the normalization condition
and we find
\begin{align}
  p(x)
  =3 \frac{(x - x_0)^2}{(x_\text{max} - x_0)^3} \; .
\end{align}
If we vary $x_0$ and $x_\text{max}$ we can trace two contributions to the
uncertainty in the density,
\begin{align}
\sigma_\text{pred} \equiv \Delta p
&\supset \frac{9}{(x_\text{max} - x_0)^4} \left| (x - x_0) \left( x - \frac{x_0}{3} - \frac{2 x_\text{max}}{3} \right) \right| \Delta x_0 \notag \\
\text{and} \qquad
\sigma_\text{pred} \equiv  \Delta p
&\supset \frac{9}{(x_\text{max} - x_0)^4} \; (x - x_0)^2 \; \Delta x_\text{max} \; ,
\label{eq:fit_kicker}
\end{align}
one from the variation of $x_0$ and one from the vriation of
$x_\text{max}$. In analogy to Eq.\eqref{eq:fit_wedge} they need to be
added in quadrature.  If the uncertainty on $\Delta x_0$ dominates,
the uncertainty has a trivial minimum at $x=0$ and a non-trivial
minimum at $x=2/3$. From $\Delta x_\text{max}$ we get another
contribution which scales like $\Delta p \propto p(x)$. In
Fig.~\ref{fig:quadratic_unc} we clearly observe both contributions,
and the green line is given by the corresponding 2-parameter fit to
the $\sigma_\text{pred}$ distribution from the BINN.

\subsection{Gaussian ring}
\label{sec:toy_ring}

\begin{figure}[b!]
\centering
\includegraphics[width=0.32\textwidth, page=1]{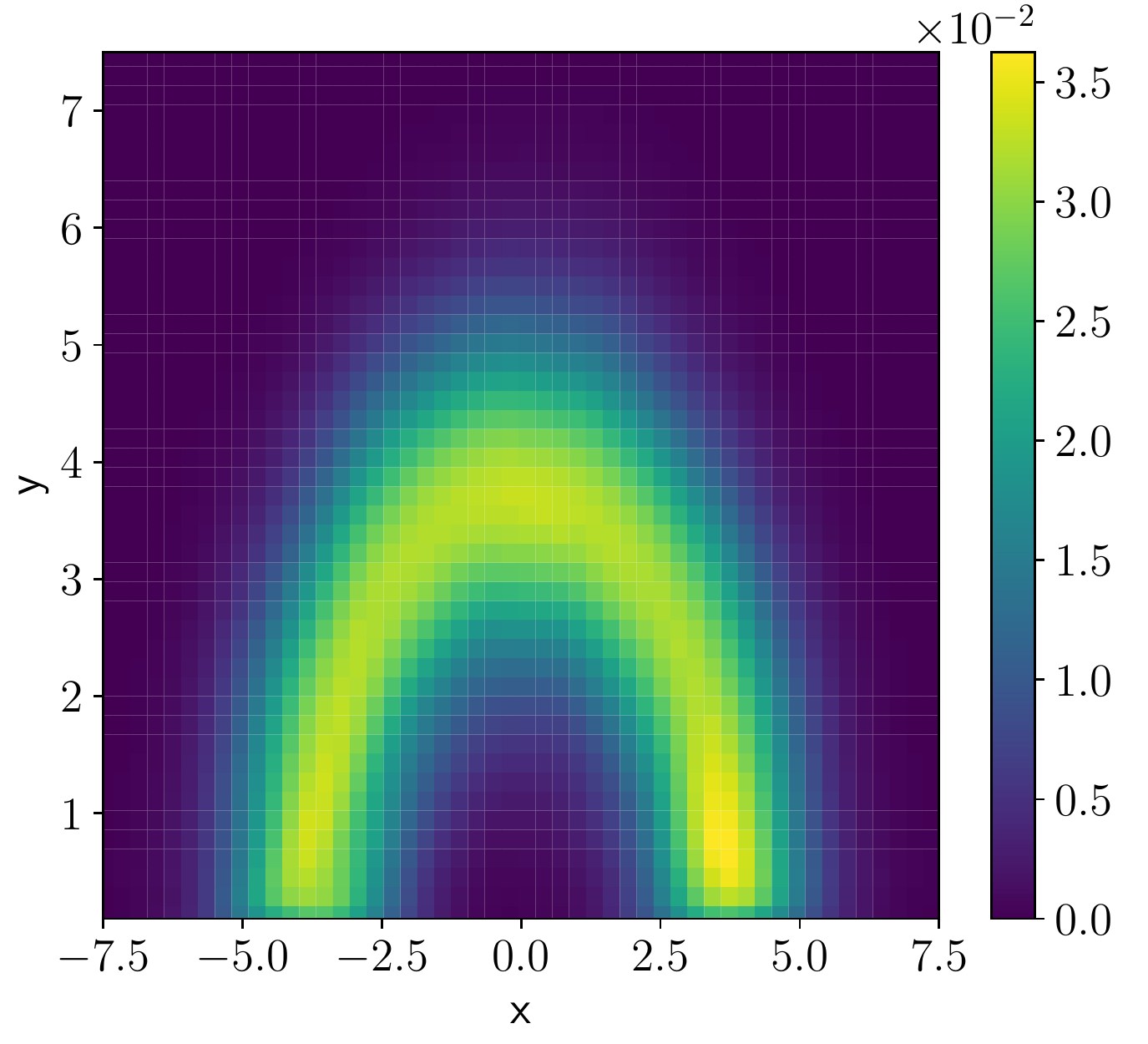}
\includegraphics[width=0.32\textwidth, page=1]{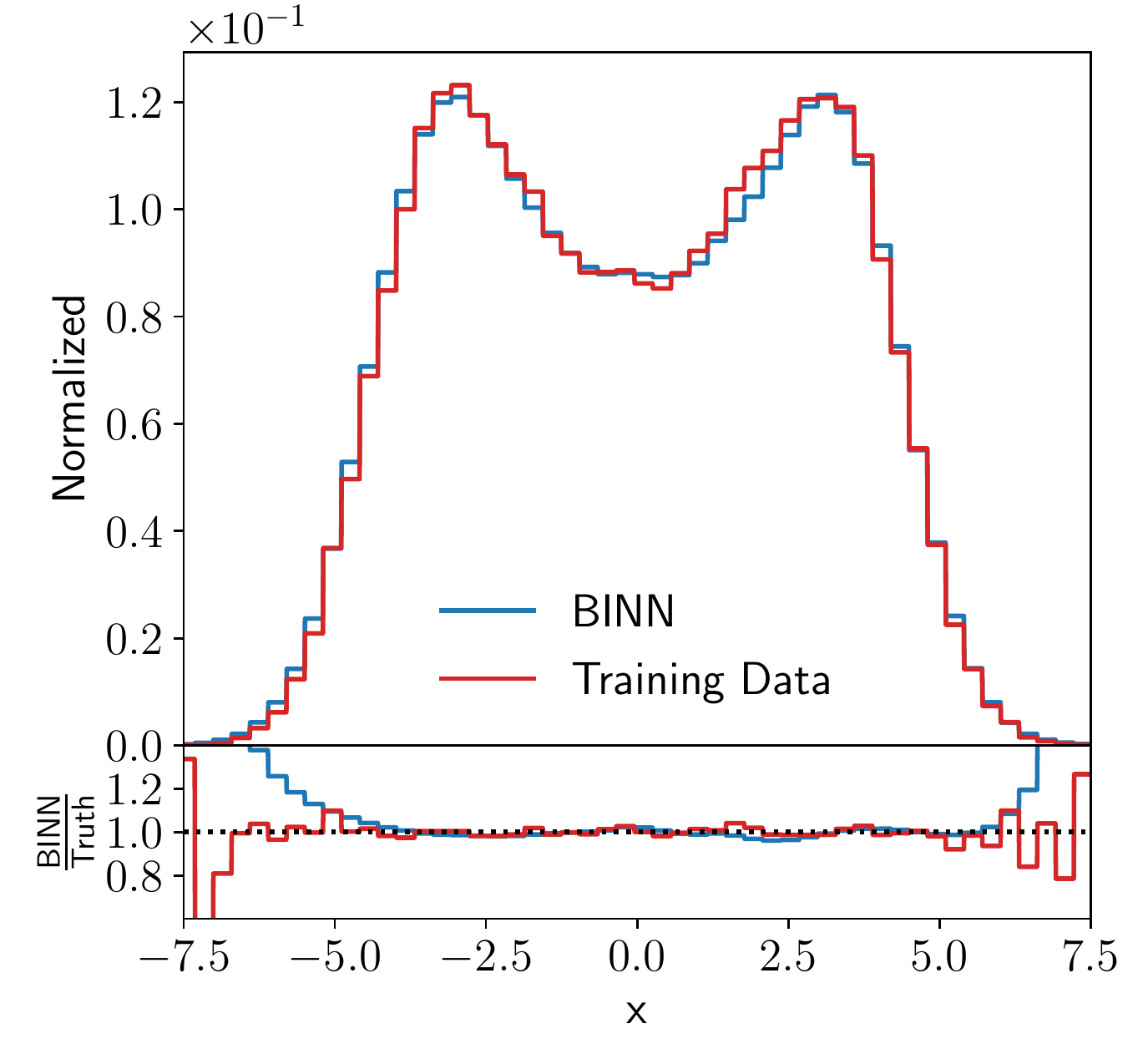}
\includegraphics[width=0.32\textwidth, page=2]{./figs/new_figures/gauss_ring_1dhists}
\caption{Two-dimensional and marginal densities for the Gaussian
  (half-)ring.}
\label{fig:gauss_hists}
\end{figure}

Our third example is a two dimensional Gaussian ring, which in terms
of polar coordinates reads
\begin{align}
p(r, \phi) = \text{Gauss}(r > 0; \mu=4, w=1) \times \text{Const}(\phi \in [0, \pi]) \; ,
\label{eq:gauss_dens}
\end{align}
We define the Gaussian density as the usual
\begin{align}
  \text{Gauss}(r) 
&=  \frac{1}{\sqrt{2 \pi} \; w} \exp \left[ - \frac{1}{2 w^2} (r-\mu)^2 \right]
\end{align}
The density defined in Eq.\eqref{eq:gauss_dens} can be translated into
Cartesian coordinates as
\begin{align}
p(x, y) = \text{Gauss}(r(x, y);\mu=4, w=1) \, \times \text{Const}(\phi(x, y) \in [0, \pi]) \times \dfrac{1}{r(x, y)}
\end{align}
where the additional factor $1/r$ comes from the Jacobian. We train
the BINN on Cartesian coordinates, just like in the two examples
before, and limit ourselves to $y>0$ to avoid problems induced by
learning a non-trivial topology in mapping the latent and phase
spaces.  In Fig.~\ref{fig:gauss_hists} we once again see that our
network describes the true two-dimensional density well.

\begin{figure}[t]
\centering
\includegraphics[width=0.32\textwidth, page=1]{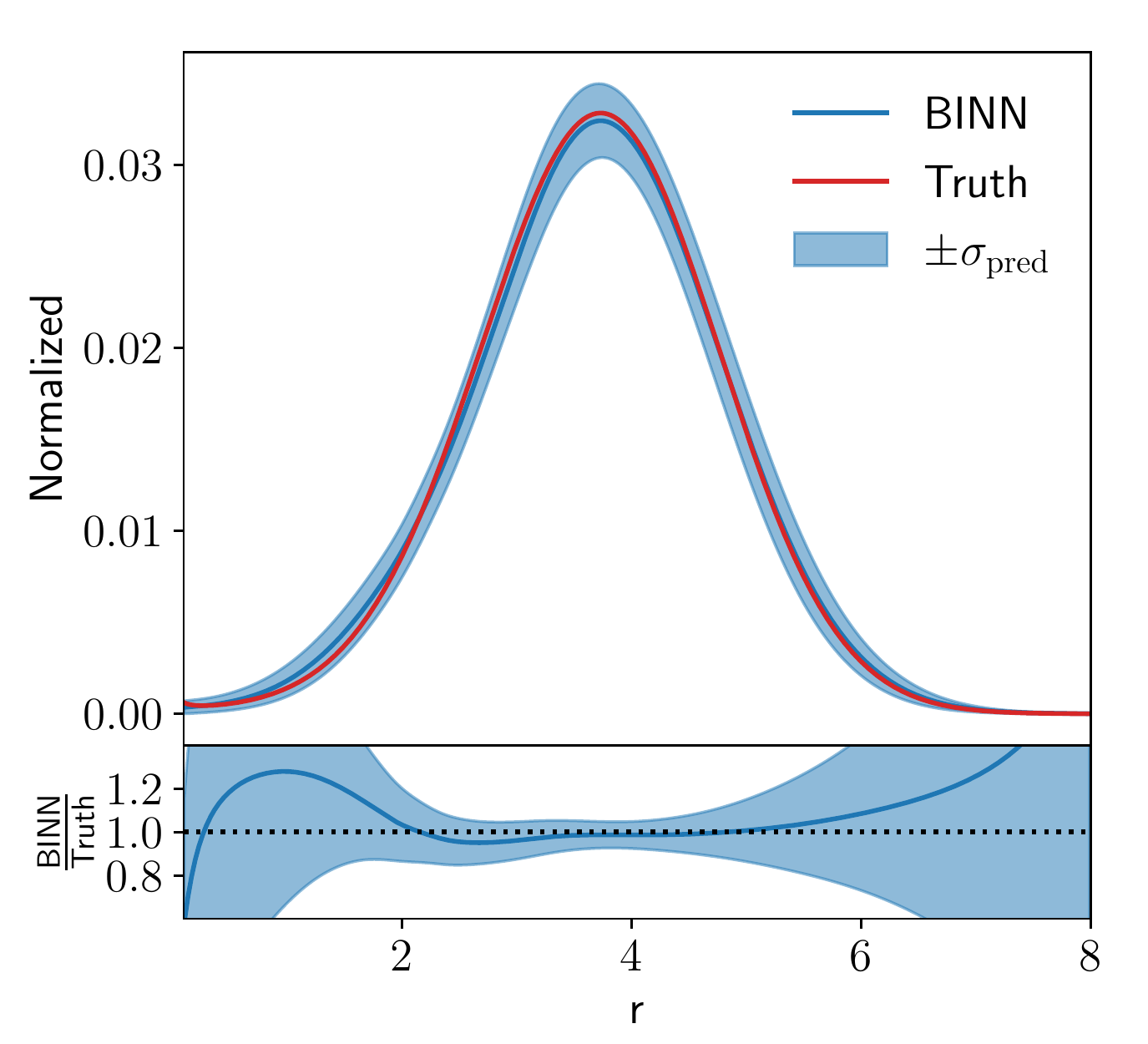}
\includegraphics[width=0.32\textwidth, page=2]{./figs/new_figures/gauss_ring_1dplots} 
\includegraphics[width=0.32\textwidth, page=3]{./figs/new_figures/gauss_ring_1dplots}
\caption{Cartesian density and predictive uncertainty distribution for
  the Gaussian ring. In the left panel the density and uncertainty are
  averaged over several lines with constant $\phi$. In the central and
  right panels, the uncertainty band on $\sigma_\text{pred}$ is given
  by their variation.  The green curve represents a two-parameter fit
  to Eq.\eqref{eq:gauss_fit}.}
\label{fig:gauss_unc}
\end{figure}

In Fig.~\ref{fig:gauss_unc} we show the Cartesian density but
evaluated on a line of constant angle. This form includes the Jacobian
and has the expected, slightly shifted peak position at $r_\text{max}
= 2 + \sqrt{3} = 3.73$. The BINN returns a predictive uncertainty,
which grows towards both boundaries.  The error band easily covers the
deviation of the density learned by the BINN and the true
density. While the relative predictive uncertainty appears to have a
simple minimum around the peak of the density, we again see that the
absolute uncertainty has a distinct structure with a local minimum
right at the peak. The question is what we can learn about the INN
from this pattern in the BINN.

As before, we describe our distribution in the relevant direction in
terms of convenient fit parameters. For the Gaussian radial density
these are the mean $\mu$ and the width $w$ used in
Eq.\eqref{eq:gauss_dens}. The contributions driven by the extraction
of the mean in Cartesian coordinates reads
\begin{align}
\sigma_\text{pred} &\equiv  \Delta p \supset
\left| \frac{G(r)}{r} \, \frac{\mu - r}{w^2} \right| \Delta \mu
\notag \\
\text{and} \qquad
\sigma_\text{pred} &\equiv \Delta p \supset
\left| \frac{(r - \mu)^2}{w^3} - \frac{1}{w} \right| \Delta w \; .
\label{eq:gauss_fit}
\end{align}
In analogy to Eq.\eqref{eq:fit_wedge} the two contributions need to be
added in quadrature for the full, fit-like uncertainty.  
The contribution from the the mean has a minimum at $r=\mu=4$ and is
otherwise dominated by the exponential behavior of the Gaussian, just
as we observe in the BINN result.
In the central and right panels we show a one-parameter fit of the BINN output and find that
the network determined the mean of the Gaussian as $\mu = 4 \pm
0.037$. We observe that including $\Delta w$ doesn't improve the goodness of the fit.


\subsection{Errors vs training statistics}
\label{sec:toy_stats}

\begin{figure}[t]
\centering
\includegraphics[width=0.32\textwidth, page=1]{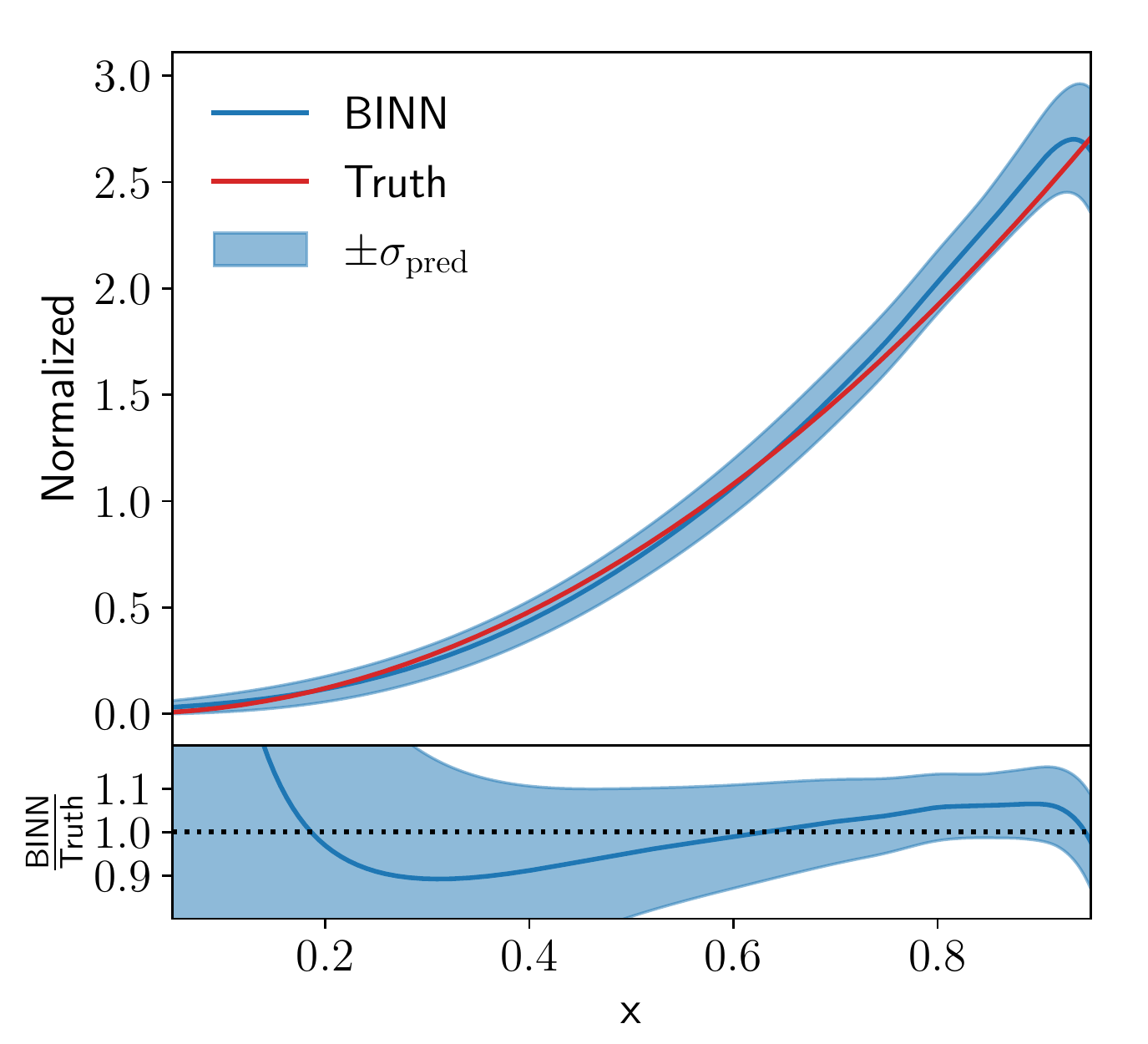}
\includegraphics[width=0.32\textwidth, page=1]{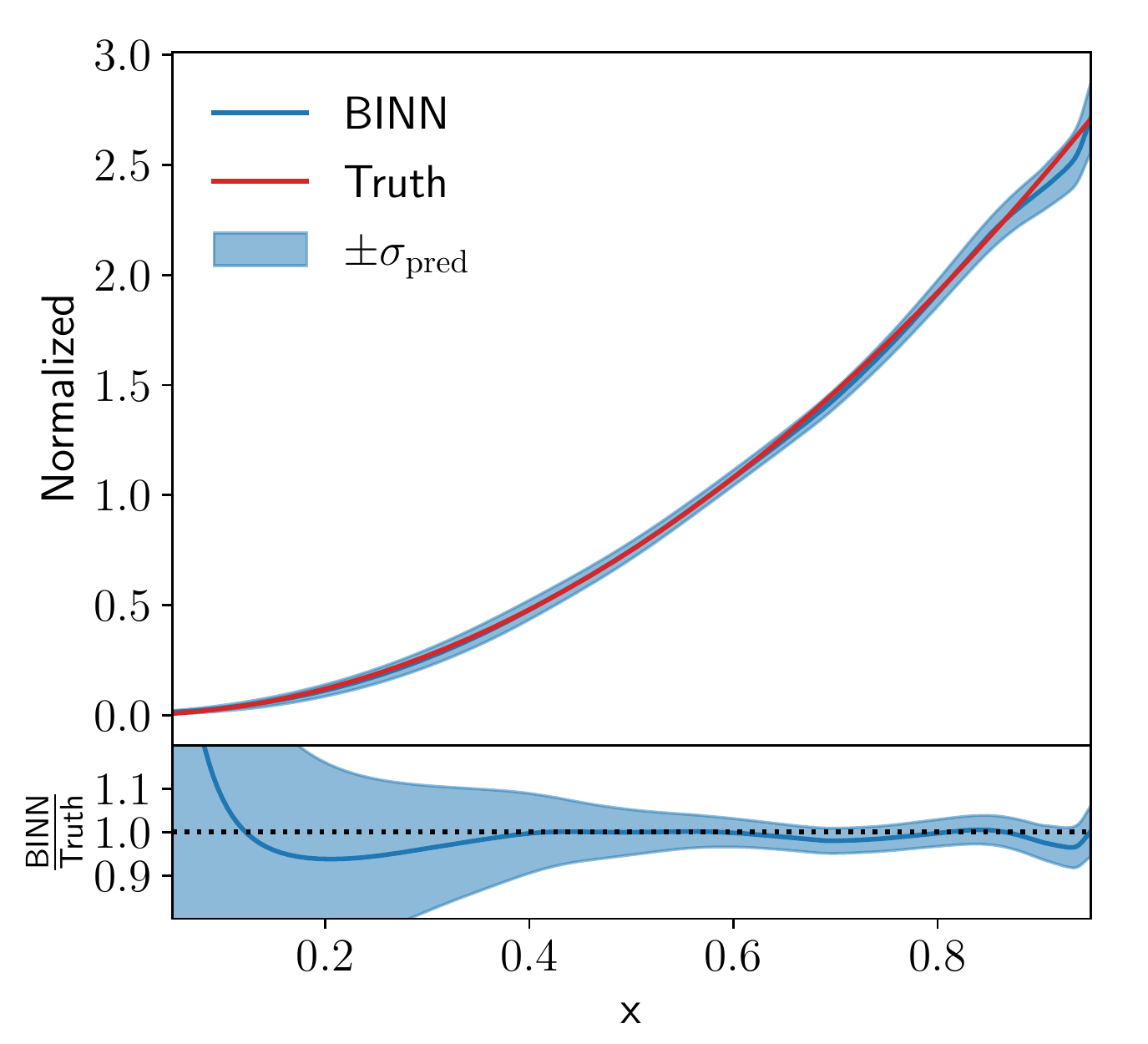}
\includegraphics[width=0.32\textwidth, page=1]{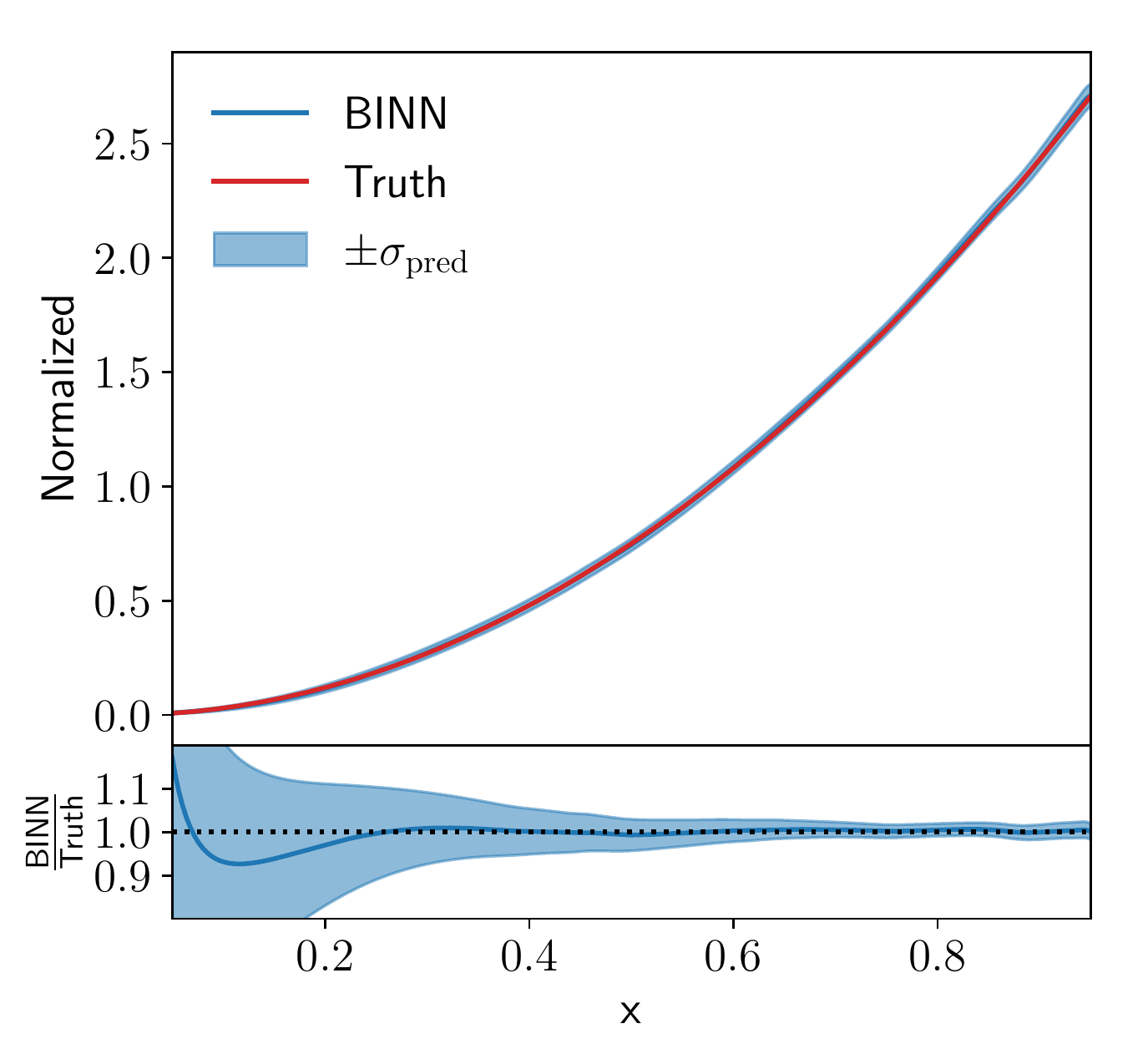} \\
\includegraphics[width=0.32\textwidth, page=2]{figs/new_figures/training_size_10k}
\includegraphics[width=0.32\textwidth, page=2]{figs/new_figures/training_size_100k} 
\includegraphics[width=0.32\textwidth, page=2]{figs/new_figures/training_size_1M} 
\caption{Dependence of the density (upper) and absolute
  uncertainty (lower) on the training statistics for the kicker ramp. We illustrate BINNs trained
  on 10k, 100k, and 1M events (left to right), to be compared to 300k
  events used for Fig.~\ref{fig:quadratic_unc}. Our training routine
  ensures that all models receive the same number of weights updates,
  regardless of the training set size.}
\label{fig:training_size}
\end{figure}

Even though it is clear from the above discussion that we cannot
expect the predictive uncertainties to have a simple scaling pattern,
like for the regression~\cite{Kasieczka:2020vlh} and
classification~\cite{Bollweg:2019skg} networks, there still remains
the question how the BINN uncertainties change with the size of the
training sample.

In Fig.~\ref{fig:training_size} we show how the BINN predictions for
the density and uncertainty change if we vary the training sample size
from 10k events to 1M training events. Note that for all toy
models, including the kicker ramp in Sec.~\ref{sec:toy_kicker}, we use
300k training events. For the small 10k training sample, we see that
the instability of the BINN density becomes visible even for our
reduced $x$-range.  The peak-dip pattern of the absolute uncertainty,
characteristic for the kicker ramp, is also hardly visible, indicating
that the network has not learned the density well enough to determine
its shape. Finally, the variation of the predictive density explodes
for $x>0.4$, confirming the picture of a poorly trained BINN. As a
rough estimate, the absolute uncertainty at $x=0.5$ with a density
value $p(x,y) = 0.75$ ranges around $\sigma_\text{pred} =
0.11~...~0.15$.

For 100k training events we see that the patterns discussed in
Sec.~\ref{sec:toy_kicker} begin to form. The density and uncertainty
encoded in the network are stable, and the peak-dip with a minimum
around $x=2/3$ becomes visible. As a rough estimate we can read off
$\sigma_\text{pred}(0.5) \approx 0.06 \pm 0.03$. For 1M training
events the picture improves even more and the network extracts a
stable uncertainty of $\sigma_\text{pred}(0.5) \approx 0.03 \pm
0.01$. Crucially, the dip around $x \approx 2/3$ remains, and even
compared to Fig.~\ref{fig:quadratic_unc} with its 300k training events
the density and uncertainty at the upper phase space boundary are much
better controlled.

Finally, we briefly comment on a frequentist interpretation of the 
BINN output. We know from simpler Bayesian networks~\cite{Bollweg:2019skg,Kasieczka:2020vlh}
that it is possible to reproduce the predictive uncertainty using 
an ensemble of deterministic networks with the same architecture. 
However, from those studies we also know that our class of Bayesian 
networks has a very efficient built-in regularization, so 
this kind of comparison is not trivial. For the BINN results shown
in this paper we find that the detailed patterns in the 
absolute uncertainties are extracted by the Bayesian network much more
efficiently than they would be for ensembles of deterministic INNs.
For naive implementations with a similar network size and no fine-tuned
regularization these patterns are somewhat harder to extract. On the 
other hand, in stable regions without distinctive patterns
the spread of ensembles of deterministic networks 
reproduces the predictive uncertainty reported by the BINN.

\subsection{Marginalizing phase space}
\label{sec:toy_marginal}

Before we move to a more LHC-related problem, we need to study how the
BINN provides uncertainties for marginalized kinematic
distribution. In all three toy examples the two-dimensional phase
space consists of one physical and one trivial direction. For
instance, the kicker ramp in Sec.~\ref{sec:toy_kicker} has a quadratic
physical direction, and in a typical phase space problem we would
integrate out the trivial, constant direction and show a
one-dimensional kinematic distribution. From our effectively
one-dimensional uncertainty extraction, we know that the absolute
uncertainty has a characteristic maximum-minimum combination, as seen
in the central panel of Fig.~\ref{fig:quadratic_unc}.

To compute the uncertainty for a properly marginalized phase space
direction, we remind ourselves how the BINN computes the density and
the predictive uncertainty by sampling over the weights,
\begin{align}
p(x, y) &= \int d\theta \, q(\theta) \, p(x, y | \theta)  \notag \\
\sigma_\text{pred}^2(x, y) &= \int d\theta \, q(\theta) \left[ p(x, y | \theta) - p(x, y) \right]^2 \, .
\label{eq:sigma_pred}
\end{align}
If we integrate over the $y$-direction, the marginalized density is
defined as
\begin{align}
  p(x)  = \int dy \, p(x,y)
       =& \int dy d\theta \, q(\theta) \, p(x, y | \theta)  \notag \\
       =& \int d\theta \, q(\theta) \, \int dy \, p(x, y | \theta)
       \equiv \int d\theta \, q(\theta) \, p(x | \theta) \; ,
\label{eq:p_marginal}
\end{align}
which implicitly defines $p(x|\theta)$ in the last step, notably
without providing us with a way to extract it in a closed form. The
key step in this definition is that we exchange the order of the $y$
and $\theta$ integrations. Nevertheless, with this definition at hand,
we can \textsl{define} the uncertainty on the marginalized
distribution as
\begin{align}
  \sigma_\text{pred}^2 (x) = \int d\theta \, q(\theta) \left[ p(x | \theta) - p(x) \right]^2 \; .
\label{eq:sigma_pred_marg}
\end{align}
We illustrate this construction with a trivial $p(x,y) = p(x,y_0)$,
where we can replace the trivial $y$-dependence by a fixed choice
$y=y_0$ just like for the wedge and kicker ramps. Here we find, modulo
a normalization constant in the $y$-integration
\begin{align}
  \sigma_\text{pred}^2 (x)
  &= \int d\theta \, q(\theta) \left[ p(x | \theta) - p(x) \right]^2
  \notag \\
  &= \int d\theta \, q(\theta) \int dy \; \left[ p(x,y_0 | \theta) - p(x,y_0) \right]^2
  \notag \\
  &= \int dy d\theta \, q(\theta) \; \left[ p(x,y_0 | \theta) - p(x,y_0) \right]^2
   = \int dy \, \sigma_\text{pred}^2 (x,y_0) = \sigma_\text{pred}^2 (x,y_0) \; .
\end{align}
Adding a trivial $y$-direction does not affect the predictive
uncertainty in the physical $x$-direction.

\begin{figure}[t]
\centering
\includegraphics[width=0.32\textwidth,page=3]{figs/new_figures/quadratic_1dhists_with_unc}
\hspace*{0.1\textwidth}
\includegraphics[width=0.32\textwidth,page=4]{figs/new_figures/quadratic_1dhists_with_unc}\\
\includegraphics[width=0.32\textwidth,page=1]{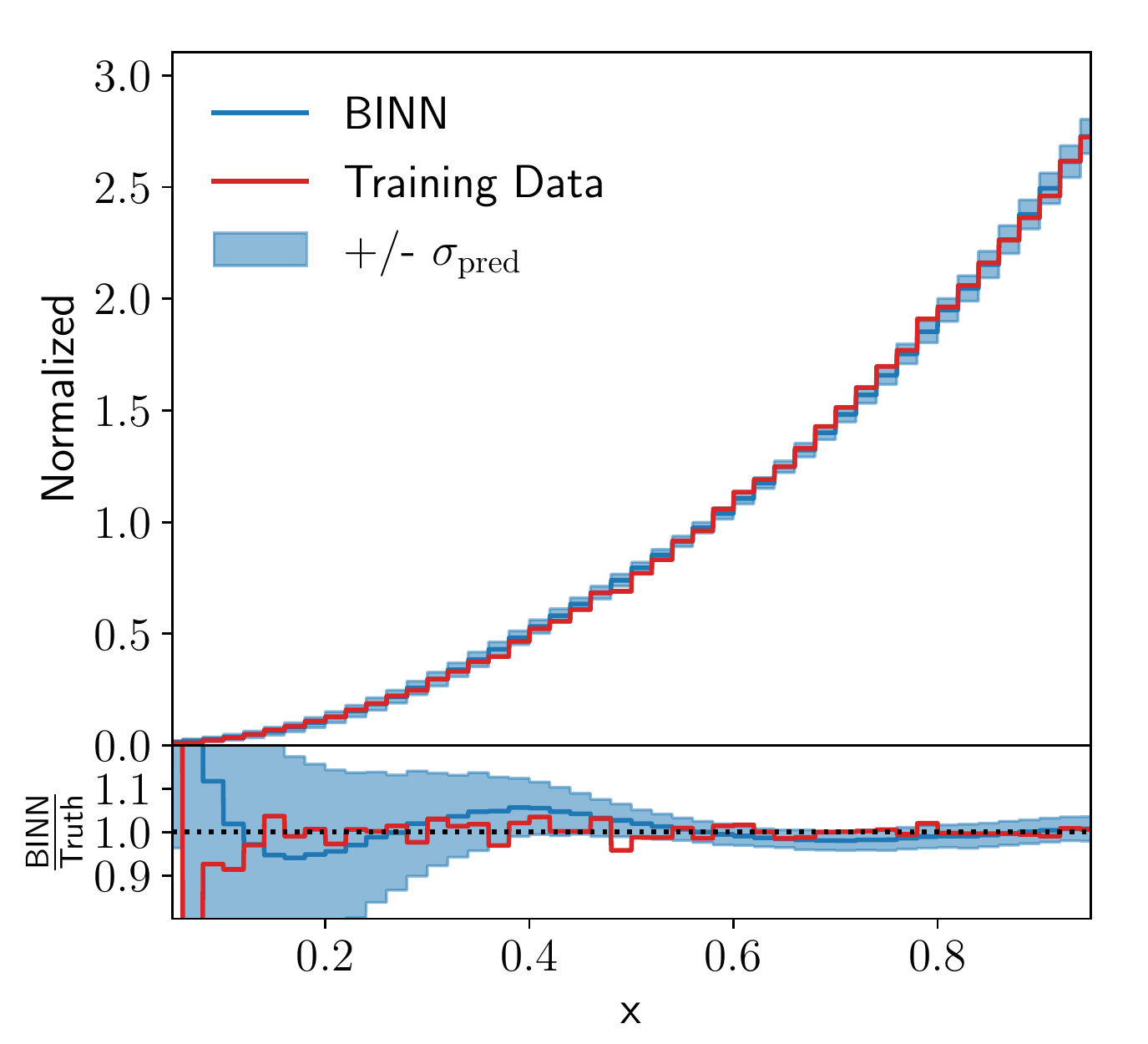}
\hspace*{0.1\textwidth}
\includegraphics[width=0.32\textwidth,page=2]{figs/new_figures/quadratic_1dhists_with_unc}
\caption{Marginalized densities and predictive uncertainties for the
  kicker ramp. Instead of the true distribution we now show the training data as a reference, to illustrate possible limitations. We use 10M phase space point to guarantee a stable prediction.}
\label{fig:marginalized}
\end{figure}

As mentioned above, unlike for the joint density $p(x, y | \theta)$,
we do not know the closed form of the marginal distributions $p(x)$ or
$p(x| \theta)$. Instead, we can approximate the marginalized
uncertainties through a combined sampling in $y$ and $\theta$.  We
start with one set of weights $\theta_i$ from the weight
distribution, based on one random number per INN weight. We now
sample $N$ points in the latent space, $z_j$, and compute $N$ phase
space points $x_j$ using the BINN configuration $\theta_i$. We then bin
the wanted phase space direction $x$ and approximate $p(x|\theta_i)$
by a histogram. We repeat this procedure $i=1~...~M$ times to extract
$M$ histograms with identical binning. This allows us to compute a
mean and a standard deviation from $M$ histograms to approximate
$p(x)$ and $\sigma_\text{pred}(x)$. The approximation of
$\sigma_\text{pred}$ should be an over-estimate, because it includes
the statistical uncertainty related to a finite number of samples per
bin.  For $N \gg 1$ this contribution should become negligible. With
this procedure we effectively sample $N \times M$ points in phase
space.

Following Eq.\eqref{eq:p_marginal}, we can also fix the phase space
points, so instead of sampling for each weight sample another set of
phase space points, we use the same phase space points for each weight
sampling. This should stabilize the statistical fluctuations, but with
the drawback of relying only on an effective number of $N$ phase space
points. Both approaches lead to the same $\sigma_\text{pred}$ for
sufficiently large $N$, which we typically set to $10^5~...~10^6$. For
the Bayesian weights we find stable results for $M=30~...~50$.

In Fig.~\ref{fig:marginalized} we show the marginalized densities and
predictive uncertainties for the kicker ramp.  In $y$-direction the
density and the predictive uncertainty show the expected flat
behavior. The only exceptions are the phase space boundaries, where the
density starts to deviate slightly from the training data and the
uncertainty correctly reflects that instability.  In $x$-direction,
the marginalized density and uncertainty can be compared to their
one-dimensional counterparts in Fig.\ref{fig:quadratic_unc}. While we
expect the same peak-dip structure, the key question is if the
numerical values for $\sigma_\text{pred}(x)$ change. If the network
learns the $y$-direction as uncorrelated additional data, the
marginalized uncertainty should decrease through a larger effective
training sample. This is what we typically see for Monte Carlo
simulations, where a combination of bins in an unobserved direction
leads to the usual reduced statistical uncertainty. On the other hand,
if the network learns that the $y$-directions is flat, then adding
events in this direction will have no effect on the uncertainty of the
marginalized distribution. This would correspond to a set of fully
correlated bins, where a combination will not lead to any improvement
in the uncertainty. In Fig.~\ref{fig:marginalized} we see that the
$\sigma_\text{pred}(x)$ values on the peak, in the dip, and to the
upper end of the phase space boundary hardly change from the
one-dimensional results in Fig.\ref{fig:quadratic_unc}. This confirms
our general observation, that the (B)INN learns a functional form of
the density in both directions, in close analogy to a fit. It also
means that the uncertainty from the generative network training is not
described by the simple statistical scaling we observed for simpler
networks~\cite{Bollweg:2019skg,Kasieczka:2020vlh} and instead points
towards a GANplification-like~\cite{Butter:2020qhk} pattern.

\section{LHC events with uncertainties}
\label{sec:lhc}

\begin{table}[b!]
\begin{small} \begin{center}
\begin{tabular}{l r }
\toprule
Parameter & Flow \\
\midrule
Hidden layers (per block) & 2\\
Units per hidden layer & 64\\
Batch size & 512 \\
Epochs & 500 \\
Trainable weights & $\sim$ 182k \\
Number of training events & $\sim$ 1M\\
Optimizer & Adam\\
($\alpha$, $\beta_1$, $\beta_2$)  & ($1\times10^{-3}$, 0.9, 0.999) \\
Coupling layers & 20 \\
Prior width & 1 \\
\bottomrule
\end{tabular}
\end{center} \end{small}
\caption{Hyper-parameters for the Drell-Yan data set, implemented in
  \pytorch(v1.4.0)~\cite{pytorch}.}
\label{tab:DY_param_details}
\end{table}

As a physics example we consider the Drell-Yan process
\begin{align}
pp \rightarrow Z \rightarrow e^+ e^- \; ,
\end{align}
with its simple $2 \to 2$ phase space combined with the parton
density. The training set consists of an unweighted set of 4-vectors
simulated with \madgraph~\cite{madgraph} at 13~TeV collider energy
with the NNPDF2.3 parton densities~\cite{Ball:2013hta}. We fix the
masses of the final-state leptons and enforce momentum conservation in
the transverse direction, which leaves us with a four-dimensional
phase space. In our discussion we limit ourselves to a sufficiently
large set of one-dimensional distributions. For these marginalized
uncertainties we follow the procedure laid out in
Sec.~\ref{sec:toy_marginal} with 50 samples in the BINN-weight
space. In Tab.~\ref{tab:DY_param_details} we give the relevant
hyper-parameters for this section.

\begin{figure}[t]
\includegraphics[width=0.32\textwidth, page=1]{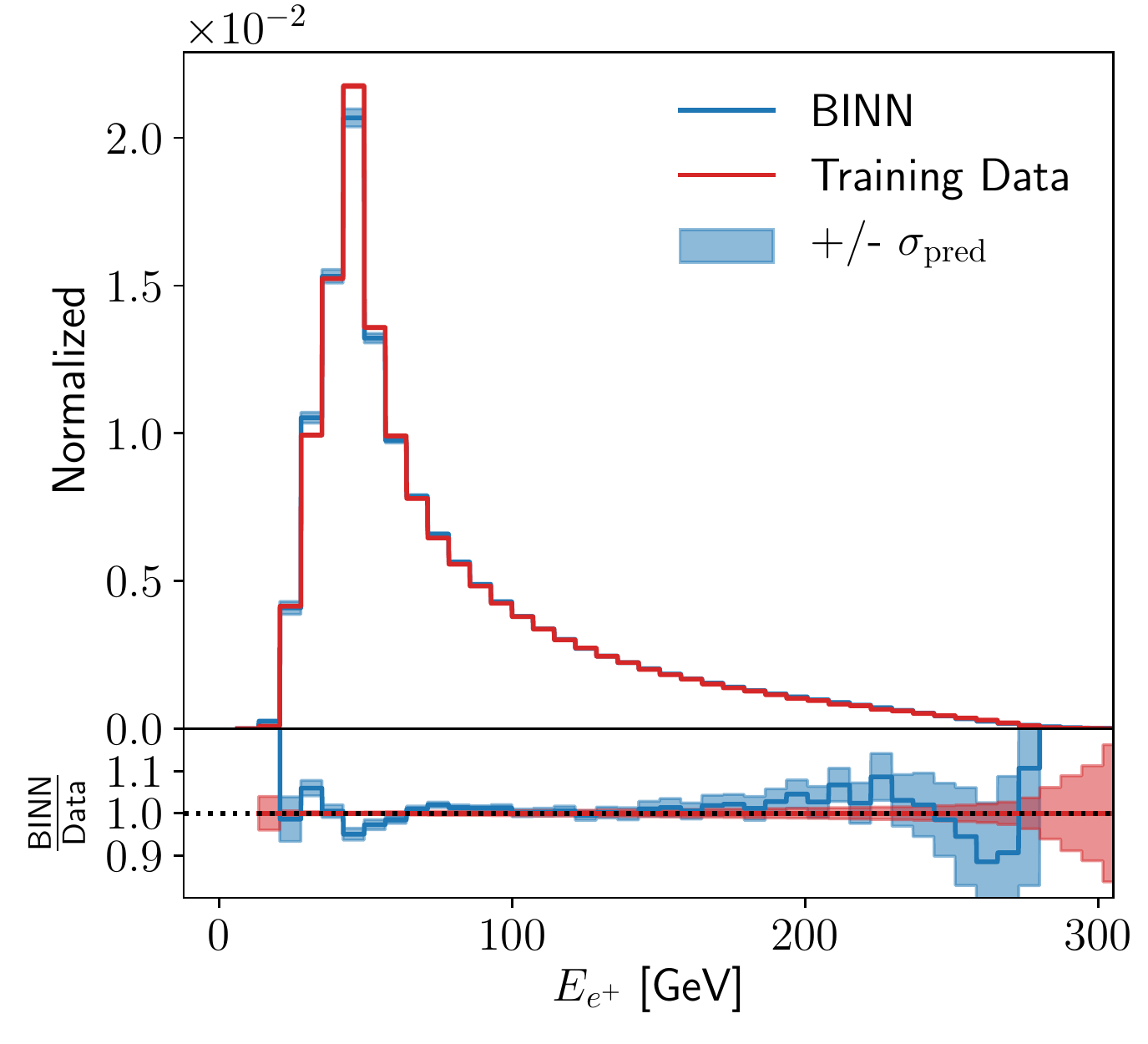}
\includegraphics[width=0.32\textwidth, page=3]{figs/new_figures/DrellYan_without_mmd_1dhists}
\includegraphics[width=0.32\textwidth, page=21]{figs/new_figures/DrellYan_without_mmd_1dhists}\\
\includegraphics[width=0.32\textwidth, page=7]{figs/new_figures/DrellYan_without_mmd_1dhists}
\includegraphics[width=0.32\textwidth, page=25]{figs/new_figures/DrellYan_without_mmd_1dhists}
\includegraphics[width=0.32\textwidth, page=27]{figs/new_figures/DrellYan_without_mmd_1dhists}
\caption{One-dimensional (marginalized) kinematic distributions for
  the Drell-Yan process.  We show the central prediction from the BINN
  and include the predictive uncertainty from the BINN as the blue
  band. The red band indicates the statistical uncertainty of the
  training data per bin in the Gaussian limit.}
\label{fig:DY}
\end{figure}

\begin{figure}[t]
\includegraphics[width=0.32\textwidth, page=2]{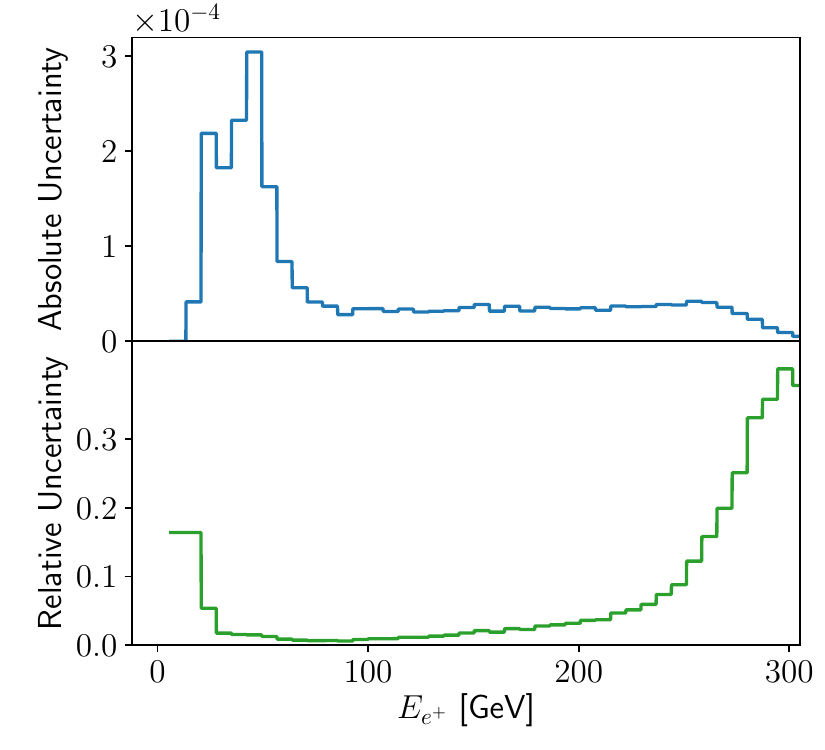}
\includegraphics[width=0.32\textwidth, page=4]{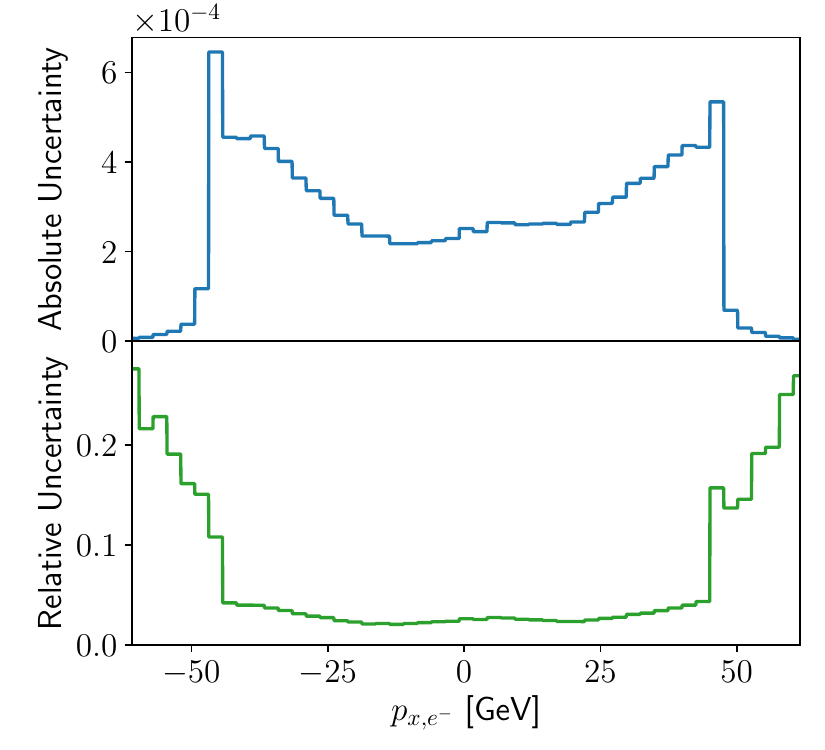}
\includegraphics[width=0.32\textwidth, page=22]{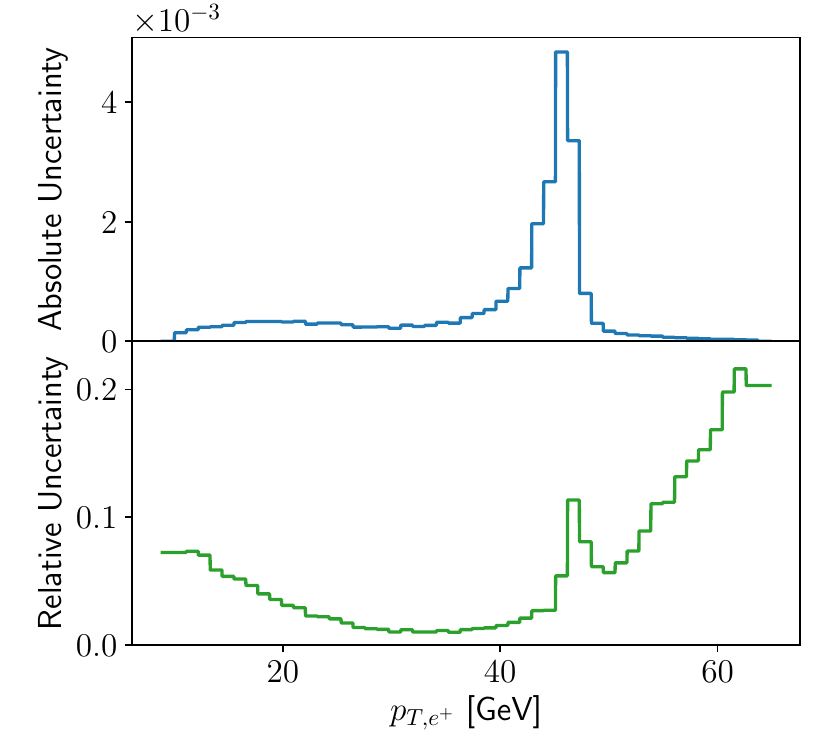}\\
\includegraphics[width=0.32\textwidth, page=8]{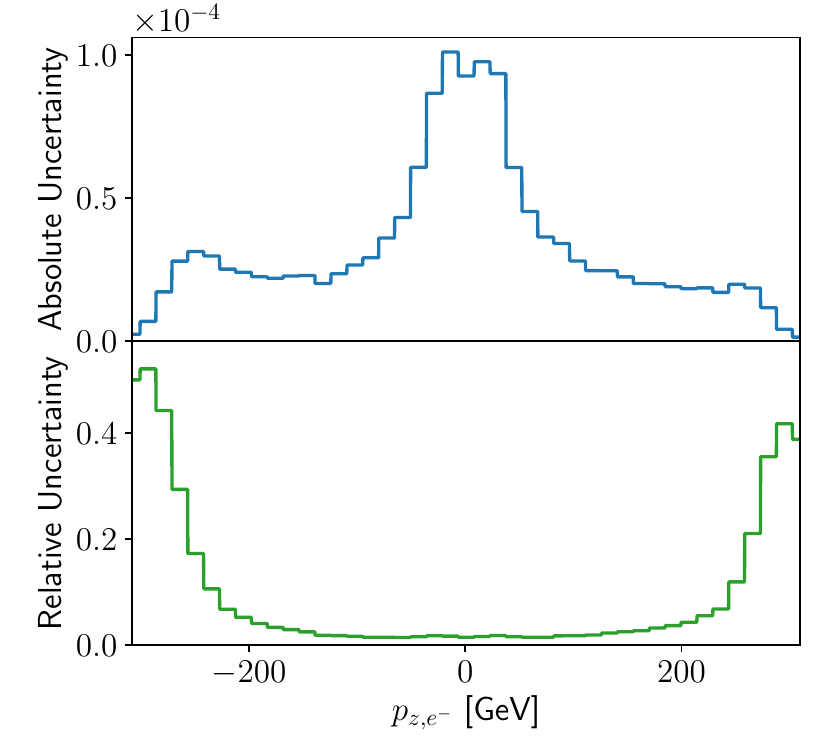}
\includegraphics[width=0.32\textwidth, page=26]{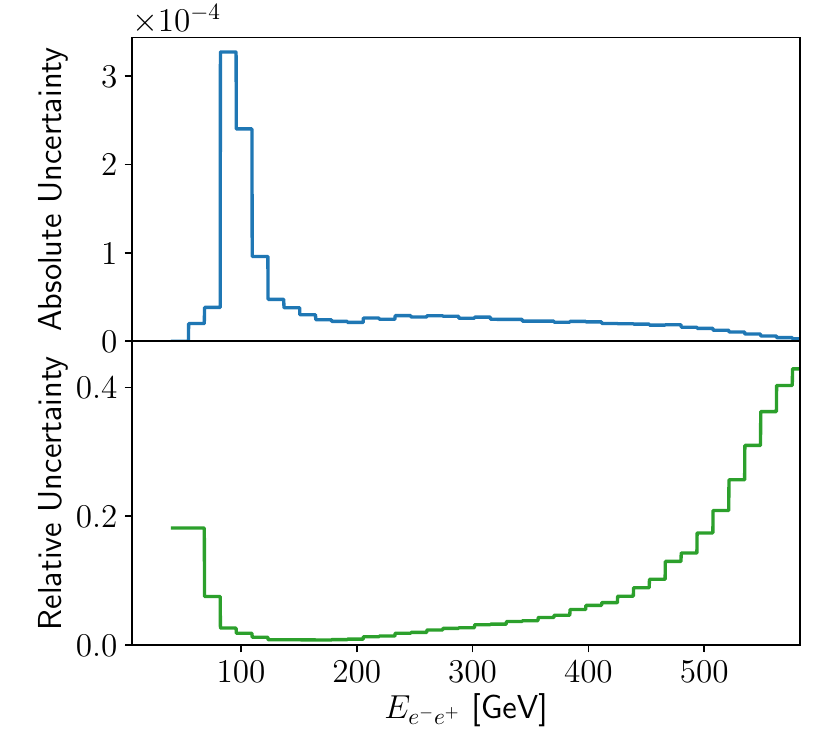}
\includegraphics[width=0.32\textwidth, page=28]{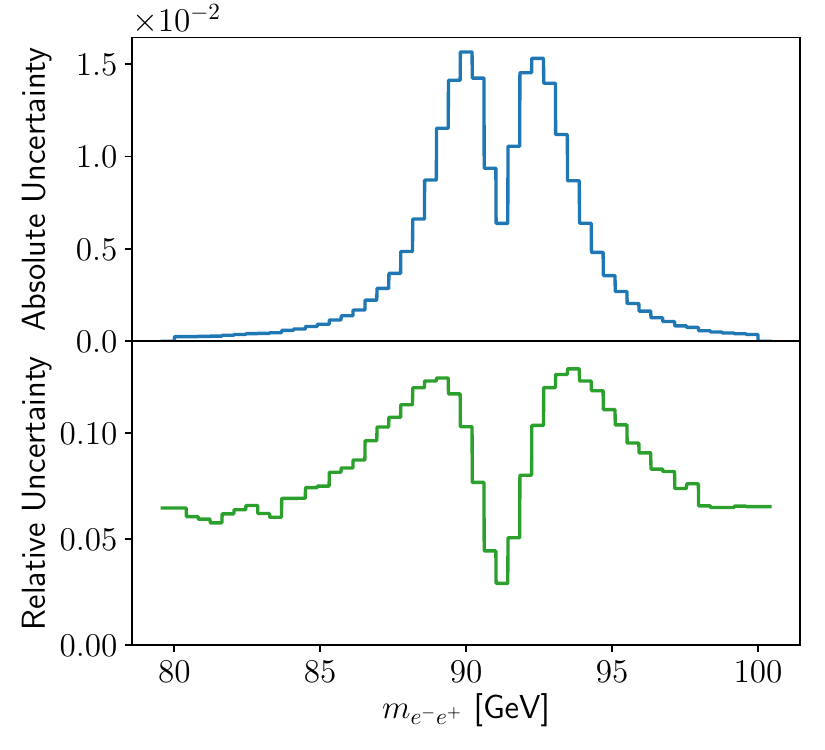}
\caption{Absolute and relative uncertainties as a function of some of
  the kinematic Drell-Yan observables shown in Fig.~\ref{fig:DY}.}
\label{fig:DY_unc}
\end{figure}

To start with, we show a set of generated kinematic distributions in
Fig.~\ref{fig:DY}. The positron energy features the expected strong
peak from the $Z$-resonance. Its sizeable tail to larger energies is
well described by the training data to $E_e \approx 280$~GeV. The
central value learned by the BINN becomes unstable at slightly lower
values of 250~GeV, as expected. The momentum component $p_x$ is not
observable given the azimuthal symmetry of the detector, but it's
broad distribution is nevertheless reproduced correctly. The
predictive uncertainty covers the slight deviations over the entire
range. What is observable at the LHC is the transverse momentum of the
outgoing leptons, with a similar distribution as the energy, just with
the $Z$-mass peak at the upper end of the distribution. Again, the
predictive uncertainty determined by the BINN covers the slight
deviations from the truth on the pole and in both tails. In the second
row we show the $p_z$ component as an example of a strongly peaked
distribution, similar to the Gaussian toy model in
Sec.~\ref{sec:toy_ring}.

While the energy of the lepton pair has a similar basic form as the
individual energies, we also show the invariant mass of the
electron-positron pair, which is described by the usual Breit-Wigner
peak. It is well known that this intermediate resonance is especially
hard to learn for a network, because it forms a narrow, highly
correlated phase space structure. Going beyond the precision shown
here would for instance require an additional MMD loss, as described
in Ref.~\cite{gan_phasespace} and in more detail in Ref.~\cite{fcgan}.
This resonance peak is the only distribution where the predictive
uncertainty does not cover the deviation of the BINN density from the
truth. This apparent failure corresponds to the fact that generative
networks always overestimate the width and hence underestimate the
height of this mass peak~\cite{gan_phasespace}.  This is an example of
the network being limited by the expressive power in phase space
resolution, generating an uncertainty which the Bayesian version
cannot account for.

In Fig.~\ref{fig:DY_unc} we show a set of absolute and relative
uncertainties from the BINN. The strong peak combined with a narrow
tail in the $E_e$ distribution shows two interesting features. Just
above the peak the absolute uncertainty drops more rapidly than
expected, a feature shared by the wedge and kicker ramps at their
respective upper phase space boundaries. The shoulder around $E_e
\approx 280$~GeV indicates that for a while the predictive uncertainty
follows the increasingly poor modelling of the phase space density by
the BINN, to a point where the network stops following the truth curve
altogether and the predictive uncertainty is limited by the expressive
power of the network.  Unlike the absolute uncertainty, the relative
uncertainty keeps growing for increasing values of $E_e$.  This
behavior illustrates that in phase space regions where the BINN starts
failing altogether, we cannot trust the predictive uncertainty either,
but we see a pattern in the intermediate phase space regime where the
network starts failing.

The second kinematic quantity we select is the (unobservable)
$x$-component of the momentum. It forms a relative flat central
plateau with sharp cliffs at each side. Any network will have trouble
learning the exact shape of such sharp phase space patterns. Here the
BINN keeps track of this, the absolute and the relative predictive
uncertainties indeed explode. The only difference between the two is
that the (learned) density at the foot of the plateau drops even
faster than the learned absolute uncertainty, so their ratio keeps
growing.

Finally, we show the result for the Breit-Wigner mass peak, the
physical counterpart of the Gaussian ring model of
Sec.~\ref{sec:toy_ring}. Indeed, we see exactly the same pattern,
namely a distinctive minimum in the predictive uncertainty right on
the mass peak.  This pattern can be explained by the network learning
the general form of a mass peak and then adjusting the mean and the
width of this peak. Learning the peak position leads to a minimum of
the uncertainty right at the peak, and learning the width brings up
two maxima on the shoulders of the mass peak.  In combination
Fig.~\ref{fig:DY} and \ref{fig:DY_unc} clearly show that the BINN
traces uncertainties in generated LHC events just as for the toy
models. Again, some distinctive patterns in the predictive uncertainty
can be explained by the way the network learns the phase space
density.

\section{Outlook}

Controlling the output of generative networks and quantifying their
uncertainties is the main task for any application in LHC physics, be
it in forward generation, inversion, or inference.  We have proposed
to use a Bayesian invertible network (BINN) to quantify the
uncertainties from the network training for each generated event. For
a series of two-dimensional toy models and an LHC-inspired
application we have shown how the Bayesian setup indeed generates an
uncertainty distribution, over the full phase space and over marginalized phase
spaces. As expected, the learned uncertainty shrinks with an improved training
statistics. Our method can be trivially extended
from unweighted to weighted events by adapting the simple MLE loss.

An intriguing result from our study is that the combined learning of
the density and uncertainty distributions allows us to draw conclusions on
how a normalizing-flow network like the BINN learns a distribution. We
find that the uncertainty distributions are naturally explained by a
fit-like behavior of the network, rather than a patch-wise learning of
the density. For the LHC, this can be seen for instance in the
non-trivial uncertainty for an intermediate Breit-Wigner
resonance. These results are another step in understanding
GANplification patterns~\cite{Butter:2020qhk} and might even allow us
to use INNs to extrapolate in phase space.

Obviously, it remains to be seen how our observations generalize to
other generative networks architectures. For the LHC, the next step
should be an in-depth study of INN-like networks applied to event
generation.

\begin{center} \textbf{Acknowledgments} \end{center}

We are very grateful for many discussions with Lynton Ardizzone, Anja
Butter, Gregor Kasieczka, Ullrich K\"othe, and Ramon Winterhalder.
The research of TP is supported by the Deutsche Forschungsgemeinschaft
(DFG, German Research Foundation) under grant 396021762 -- TRR~257
\textsl{Particle Physics Phenomenology after the Higgs Discovery}. MB
is supported by the International Max Planck School \textsl{Precision
  Tests of Fundamental Symmetries}. ML is supported by the DFG
Research Training Group GK-1940, \textsl{Particle Physics Beyond the
  Standard Model}.

\bibliography{literature}
\end{document}